\PassOptionsToPackage{table}{xcolor}
\documentclass{svmult}       
\smartqed  
\usepackage{xcolor}
\usepackage{graphicx}
\usepackage{proof}
\usepackage{amsmath}
\usepackage{tikz}
\usepackage{cmll}
\usetikzlibrary{arrows,matrix,graphs,shapes,snakes,automata,backgrounds,petri,positioning,calc,decorations.pathmorphing}
\usepackage{amssymb}
\usepackage{stmaryrd}
\usepackage[backend=biber,style=authoryear-icomp]{biblatex}
\newcommand{\one}{\mathbf{1}}
\newcommand{\editout}[1]{}
\newcommand{\defend}{\hfill $\blacklozenge$}
\newcommand{\poslin}[1]{[ #1 ]}
\newcommand{\bigpossep}{\;}
\newcommand{\possep}{\,}
\newcommand{\bo}{}
\newcommand{\bc}{}

\newcommand{\lolli}{\multimap}

\newcommand{\incomparable}{\mathbin{|}}

\newcommand{\coverseq}{\mathbin{\preceq}}

\newcommand{\onefree}{open}

\newtheorem{constraint}{Constraint}

\tikzstyle{reverseclip}=[insert path={(current page.north east) --
  (current page.south east) --
  (current page.south west) --
  (current page.north west) --
  (current page.north east)}
]
\tikzset{invclip/.style={clip,insert path={{[reset cm]
      (-16383.99999pt,-16383.99999pt) rectangle (16383.99999pt,16383.99999pt)
    }}}}

\tikzset{pas/.style={fill=gray!60}, 
act/.style={fill=gray!30},
main/.style={draw,fill=white},
ctx/.style={rounded rectangle,minimum size=7mm},
val/.style={rectangle,minimum size=7mm},
cmd/.style={chamfered rectangle,draw,fill=white},
tns/.style={circle,minimum size=4mm,draw,fill=white},
par/.style={circle,minimum size=4mm,draw,fill=black}, 
minipar/.style={circle,minimum size=2.5mm,draw,fill=black}, 
pn/.style={rounded corners, rectangle,fill=blue!30,draw,minimum size=15mm},
medpn/.style={rounded corners, rectangle,fill=blue!30,draw,minimum size=20mm},
 bigpn/.style={rounded corners, rectangle,fill=blue!30,draw,minimum size=25mm}}

\begin{document}

\title{Partial Orders, Residuation, and First-Order Linear Logic
}
%

\author{Richard Moot}


\institute{Richard Moot\at
              Universit\'{e} de Montpellier, LIRMM, CNRS\\
              161 rue Ada
              34095 Montpellier Cedex 5\\
              France\\
              Tel.: +33-4-67418585\\
              Fax: +33-4-67418500\\
              \email{firstname.lastname@institute.country}           
}

\date{Received: date / Accepted: date}

\maketitle

\abstract{
We will investigate proof-theoretic and linguistic aspects of first-order linear logic. We will show that adding partial order constraints in such a way that each sequent defines a unique linear order on the antecedent formulas of a sequent allows us to define many useful logical operators. In addition, the partial order constraints improve the efficiency of proof search. 
}

\section{Introduction}
\label{sec:intro}


Residuation is a standard principle which holds for the Lambek calculus and many of its variants. However, even though first-order linear logic can embed the Lambek calculus and some of its variants, linear logic formulas need not be part of a residuated triple (or pair). In this paper, we will present conditions on first-order linear logic in the form of partial order constraints which allow it to satisfy the residuation principle. We investigate the number of connectives definable this way and compare these connectives to the connectives definable in other type-logical grammars. We conclude by investigating some of the applications of these results, both in terms of linguistic modelling and in terms of improving upon the efficiency of proof search.

\section{Categorial Grammars and Residuation}
\label{sec:cgres}

Lambek introduced his syntactic calculus first as a calculus based on residuation \parencite[Section~7, with a sequent calculus in Section~8]{lambek}. The principle of residuation is shown as Equation~\ref{eq:res}. 

\begin{equation}\label{eq:res}
A \rightarrow C\mathbin{/} B \quad\Longleftrightarrow\quad A\mathbin{\bullet}B\rightarrow C \quad\Longleftrightarrow\quad B\rightarrow A\mathbin{\backslash} C
\end{equation}

The Lambek calculus is then defined using just the principle of residuation together together with reflexivity and transitivity of the derivation arrow and associativity of the product `$\bullet$'. Table~\ref{tab:res} lists the full set of rules of the residuation-based representation of the Lambek calculus.

\begin{table}
\begin{center}
\begin{tabular}{ccc}
\multicolumn{3}{c}{\textbf{Identity}} \\[2mm]
\infer[\textit{Refl}]{A\rightarrow A}{} && 
\infer[\textit{Trans}]{A\rightarrow C}{A\rightarrow B & B\rightarrow C} \\[2mm]
\multicolumn{3}{c}{\textbf{Residuation}} \\[2mm]
	\infer[\textit{Res}_{\bullet,/}]{A\rightarrow C\mathbin{/} B}{A\mathbin{\bullet}B\rightarrow C} && 
	\infer[\textit{Res}_{\bullet,\backslash}]{B\rightarrow A\mathbin{\backslash} C}{A\mathbin{\bullet}B\rightarrow C} \\[2mm]
	\infer[\textit{Res}_{/,\bullet}]{A\mathbin{\bullet}B\rightarrow C}{A\rightarrow C\mathbin{/} B} &&
	\infer[\textit{Res}_{\backslash,\bullet}]{A\mathbin{\bullet}B\rightarrow C}{B\rightarrow A\mathbin{\backslash} C} \\[2mm]
	\multicolumn{3}{c}{\textbf{Associativity}} \\[2mm]
\infer[\textit{Ass}_1]{A\mathbin{\bullet}( B \mathbin{\bullet} C) \rightarrow (A\mathbin{\bullet} B) \mathbin{\bullet} C}{} 	&&
\infer[\textit{Ass}_2]{(A\mathbin{\bullet} B) \mathbin{\bullet} C \rightarrow A\mathbin{\bullet}( B \mathbin{\bullet} C)}{}
\end{tabular}
\end{center}

\caption{Residuation-based presentation of the Lambek calculus}
\label{tab:res}	
\end{table}

In the Lambek calculus, the standard interpretation of the product `$\bullet$' is as a type of concatenation, with the implications `$\backslash$' and `$/$' its residuals. Using the residuation calculus, we can derive standard cancellation schemes such as the following.
\begin{align*}
\infer[\textit{Res}_{/,\bullet}]{(C\mathbin{/} B)\mathbin{\bullet} B\rightarrow C}{\infer[\textit{Refl}]{C\mathbin{/} B\rightarrow C\mathbin{/} B}{}} &&
\infer[\textit{Res}_{\backslash,\bullet}]{A \mathbin{\bullet} (A\mathbin{\backslash} C) \rightarrow C}{\infer[\textit{Refl}]{A\mathbin{\backslash} C\rightarrow A\mathbin{\backslash} C}{}}
\end{align*}

Showing us that when we compose $C\mathbin{/} B$ with a $B$ to its right, we produce a $C$, and that when we compose $A\mathbin{\backslash} C$ with an $A$ to its left, we produce a $C$.

\begin{figure}
\begin{center}
\begin{tikzpicture}
\node (c) at (4em,0em) {$A\mathbin{\bullet} B$};
\node (apb) at (4em,-1.5em) {$C$};
\node (bc) at (0em,5em) {$C\mathbin{/} B$};
\node (a) at (0em,6.5em) {$A$};
\node (b) at (8em,6.5em) {$B$};
\node (ac) at (8em,5em) {$A\mathbin{\backslash} C$};
\draw (c) -- (bc);
\draw (c) -- (ac);
\draw (ac) -- (bc); 	
\end{tikzpicture}
\end{center}
\caption{Visual representation of residuation}
\label{fig:visres}
\end{figure}
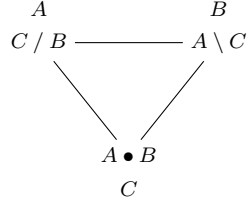

Figure~\ref{fig:visres} shows a standard visual representation of the residuation principle in the form of a triangle, with the each of the vertices of the triangle corresponding to one of the Lambek calculus connectives. 

We can `read off' many of the principles from this triangle, for example, the three different ways of concatenating the elements of a residuated triple are:
\begin{enumerate}
\item 	  composing $A$ and $B$ to produce $A\bullet B$,
\item composing $C\mathbin{/} B$ and $B$ to produce $C$,
\item composing $A$ and $A\mathbin{\backslash} C$ to produce $C$.
\end{enumerate}

The residuation presentation of the Lambek calculus naturally forms a category. This not only gives the Lambek calculus a category theoretic foundation --- something \textcite{girard11} argues is an important, deeper level of meaning for logics --- but it can also play the role of an alternative type of natural language semantics for the Lambek calculus \parencite{lambek88,coecke2013lambek}, to be contrasted with the more standard semantics for type-logical grammars in the tradition of \textcite{montague}.

\begin{table}
\begin{center}
\begin{tabular}{ccc}
\multicolumn{3}{c}{\textbf{Identity}} \\[2mm]
\infer[\textit{Refl}]{A\rightarrow A}{} && 
\infer[\textit{Trans}]{A\rightarrow C}{A\rightarrow B & B\rightarrow C} \\[2mm]
\multicolumn{3}{c}{\textbf{Application}} \\[2mm]
\infer[\textit{Appl}\backslash]{A\mathbin{\bullet}(A\mathbin{\backslash} B)\rightarrow B}{} &&
\infer[\textit{Appl}/]{(B\mathbin{/}A)\mathbin{\bullet} A\rightarrow B}{} \\[2mm]
\multicolumn{3}{c}{\textbf{Co-Application}} \\[2mm]
\infer[\textit{Coappl}\backslash]{A\rightarrow B\mathbin{\backslash}(B\mathbin{\bullet}A)}{} &&
\infer[\textit{Coappl}/]{A\rightarrow (A\mathbin{\bullet}B)\mathbin{/}B}{} \\[2mm]
\end{tabular}
\begin{tabular}{ccc}
\multicolumn{3}{c}{\textbf{Monotonicity}} \\[2mm]
	\infer[\textit{Mon}_{\backslash}]{B\mathbin{\backslash} C\rightarrow A\mathbin{\backslash} D}{A\rightarrow B& C\rightarrow D} &
	\infer[\textit{Mon}_{\bullet}]{A\mathbin{\bullet} C\rightarrow B\mathbin{\bullet} D}{A\rightarrow B & C\rightarrow D} & 
	\infer[\textit{Mon}_{/}]{C\mathbin{/}B\rightarrow D\mathbin{/}A}{A\rightarrow B & C\rightarrow D} \\[2mm]
\end{tabular}
\begin{tabular}{ccc}
\multicolumn{3}{c}{\textbf{Associativity}} \\[2mm]
\infer[\textit{Ass}_1]{A\mathbin{\bullet}( B \mathbin{\bullet} C) \rightarrow (A\mathbin{\bullet} B) \mathbin{\bullet} C}{} 	&&
\infer[\textit{Ass}_2]{(A\mathbin{\bullet} B) \mathbin{\bullet} C \rightarrow A\mathbin{\bullet}( B \mathbin{\bullet} C)}{}
\end{tabular}
\end{center}
\caption{Do\v{s}en's presentation of the Lambek calculus}
\label{tab:dosen}	
\end{table}

An alternative combinatorial representation of residuation is found in Table~\ref{tab:dosen}. This presentation uses the two application principles we have derived above as axioms, and adds two additional principles of co-application, easily obtained from the identity on the product formulas together with a residuation step. 
\begin{align*}
\infer[\textit{Res}_{\bullet,/}]{A\rightarrow (A \mathbin{\bullet} B) \mathbin{/} B}{\infer[\textit{Refl}]{A \mathbin{\bullet} B\rightarrow A \mathbin{\bullet} B}{}}&&
\infer[\textit{Res}_{\bullet,\backslash}]{A\rightarrow B\mathbin{\backslash}(B\mathbin{\bullet} A)}{\infer[\textit{Refl}]{B\mathbin{\bullet} A\rightarrow B\mathbin{\bullet} A}{}} 
\end{align*}

The advantage of this presentation is that, besides transitivity, the only recursive rules are the monotonicity principles for the three connectives. This makes this presentation especially convenient for inductive proofs. For example, the completeness proofs of \textcite{dosen92frames} use this presentation.

\subsection{Residuation in Extended Lambek Calculi}

Many of the extensions and variants of the Lambek calculus which have been proposed keep the principle of residuation central. For example, the multimodal Lambek calculus simply uses multiple families of residuated connectives $\{/_{i}, \bullet_i, \backslash_i \}$ for members $i$ of a fixed, small set $I$ of modes. Similarly, the unary connectives `$\Diamond$' and `$\Box$' connectives are a residuated pair \parencite{mmli,KurtoMM,oehrle11multi}.



However, some other formalisms do not use residuation as their central tool for defining connectives. These formalisms either add connectives corresponding to alternative algebraic principles, or abandon residuation altogether.

Formalisms in the former group take residuation for some of its connectives and \emph{add} additional principles such as dual residuation, Galois connections, and dual Galois connections for other connectives \parencite{areces2004galois,bernardi10conti}. 

Formalisms in the latter group abandon residuation as a key principle (without replacing it with another algebraic principle), or only preserve it for some of their connectives. These formalisms include lambda grammars \parencite{oehrle}, hybrid type-logical grammars \parencite{kl12gap,kl20tls} and first-order linear logic \parencite{mill1,moot13lambek}.

\subsection{Residuation and First-Order Linear Logic}

The main theme of this paper will be to investigate what types of connectives are definable in first-order linear logic when we restrict ourselves to residuated connectives. We will look at generalised forms of concatenation and their residuals and see how we can define these in first-order linear logic.


Some of these definable connectives require us to explicitly specify partial order constraints on some of the positions to preserve the required information. The resulting grammar system then has two components: for a sentence to be grammatical, a logical statement has to be derivable (as is standard for type-logical grammars) but also a corresponding partial order definition must be consistent. This gives us a mechanism to specify the relative order of grammatical constituents (logical formulas in type-logical grammars). The property we want to preserve locally in each statement is that the strings corresponding to the antecedent formulas can be linearly ordered in a unique way.  

\section{First-Order Linear Logic}
\label{sec:foll}

A \emph{sequent} or a \emph{statement} is an expression of the form $A_1,\ldots,A_n \vdash C$ (for some $n \geq 0$), which we will often shorten to $\Gamma \vdash C$. We call $\Gamma$ the \emph{antecedent}, formulas $A_i$ in $\Gamma$ \emph{antecedent formulas}, and $C$ the \emph{succedent} of the statement. We assume the sequent comma is both associative and commutative and treat statements which differ only with respect to the order of the antecedent formulas to be equal.
Table~\ref{tab:seqmill1} shows the sequent calculus rules for first-order multiplicative intuitionistic linear logic. The $R \forall$ and $L \exists$ rule have the standard side condition that there are no free occurrences of $x$ in $\Gamma$ and $C$. 

\begin{table}
\begin{center}
\begin{tabular}{cc}

\infer[\bo Ax\bc]{A \vdash A}{} &

\infer[\bo Cut\bc]{\Gamma,\Delta \vdash C}{\Gamma \vdash A & \Delta,A
\vdash C} \\[2mm]

\infer[\bo L\otimes\bc]{\Gamma,A\otimes B \vdash C}{\Gamma,A,B \vdash
C} &

\infer[\bo R\otimes\bc]{\Gamma,\Delta \vdash A\otimes B}{\Gamma \vdash
A & \Delta \vdash B} \\[2mm]

\infer[\bo L\lolli\bc]{\Gamma,\Delta,A\lolli B\vdash C}{\Delta \vdash
A & \Gamma,B \vdash C} &

\infer[\bo R\lolli\bc]{\Gamma \vdash A\lolli B}{\Gamma,A \vdash B} \\[2mm]

\infer[\bo L\exists^*\bc]{\Gamma, \exists x.A \vdash C}{\Gamma, A
\vdash C} &

\infer[\bo R\exists\bc]{\Gamma \vdash \exists x.A}{\Gamma \vdash A[x:=t]} \\[2mm]

\infer[\bo L\forall\bc]{\Gamma, \forall x.A \vdash C}{\Gamma, A[x:=t] \vdash
C} &

\infer[\bo R\forall^*\bc]{\Gamma \vdash \forall x.A}{\Gamma \vdash
A} \\

\end{tabular}
\end{center}
\caption{The sequent calculus for first-order intuitionistic multiplicative linear logic.}
\label{tab:seqmill1}
\end{table}

The sequent calculus is decidable (the decision problem is NP complete \parencite{lincoln}) and sequent proof search can be used as a practical decision procedure \parencite{foll}. Decidability presupposes both cut elimination (which, as usual, is a simple enough proof even though it consists of many rule permutation cases to verify) and a restriction on the choice of $t$ for the $L\forall$ and $R\exists$ rules. A standard solution is to use unification for this purpose, effectively delaying the choice of $t$ to the most general term required by the axioms in backward chaining cut-free proof search. This of course requires us to verify the eigenvariable conditions for the $R \forall$ and $L \exists$ rules are still satisfied after unification. We can see this in action in the following failed attempt to prove $\forall y [a \otimes b(y)] \vdash a\otimes  \forall x.b(x)$ (the reader can easily verify all other proof attempts fail as well). 
\[
\infer[L\forall]{\forall y. [a \otimes b(y)] \vdash a\otimes  \forall x.b(x)}{
    \infer[L\otimes]{a \otimes b(Y) \vdash a\otimes  \forall x.b(x)}{
        \infer[R\otimes]{a, b(Y) \vdash a\otimes  \forall x.b(x)}{
            \infer[Ax]{a\vdash a}{}
          & \infer[\forall R*]{b(Y)\vdash \forall x.b(x)}{\infer[Ax]{b(Y) \vdash b(x)}{Y = x}}
        }
    }
}
\]

Tracing the proof from the endsequent upwards to the axioms, we start by replacing $y$ by a fresh metavariable $Y$ to be unified later, then follow  the proof upwards to the axioms. For the $b$ predicates, we compute the most general unifier of $x$ and $Y$, which is $x$. But then, the antecedent of the $\forall R$ rule becomes $b(x)$, which fails to respect the eigenvariable condition for $x$. We can improve on the sequent proof procedure for first-order linear logic, even exploiting some of the rule permutabilities \parencite{foll}. However,  in Section~\ref{sec:pn} we will present a proof net calculus for first order linear logic, following \textcite{quant}, which \emph{intrinsically} avoids the efficiency problems caused by rule permutations.  

Before we do so, however, we will briefly recall how we can use first-order linear logic for modelling natural languages.
  
\subsection{First-Order Linear Logic and Natural Language Grammars}

For type-logical grammars, a \emph{lexicon} is a mapping from words to formulas in the corresponding logic. In first-order linear logic, this mapping is parametric for two position variables $L$ and $R$, corresponding respectively to the left and right position of the string segment corresponding to the word. In general, for a sentence with $n$ words, we assign the formula of word $w_i$ (for $1\leq i\leq n$) the string positions $i-1$ and $i$. This simply follows the fairly standard convention in the parsing literature to represent substrings of the input string by pairs of integers.

As noted by \textcite{mill1}, we can translate Lambek calculus formulas to first-order linear logic formulas as follows.
\begin{align}
\| p \|^{x,y} &= p(x,y) \\
\| A\mathbin{\bullet} B \|^{x,z} & = \exists y. \| A \|^{x,y} \otimes \| B \|^{y,z} \\
\| A\mathbin{\backslash} C \|^{y,z} & = \forall x. \| A \|^{x,y} \multimap \| C \|^{x,z} \\
\label{tr:dr}\| C\mathbin{/} B \|^{x,y} & = \forall z. \| B \|^{y,z} \multimap \| C \|^{x,z} 	
\end{align}

Equation~\ref{tr:dr} states that when $C/B$ is a formula spanning string $x,y$ (that is, having $x$ as its left edge and $y$ as its right edge), that means combining it with a formula $B$ having $y$ as its left edge and any $z$ as its right edge.

\begin{figure}
\begin{center}
\begin{tikzpicture}
\node (c) at (4em,0em) {$\smash{\exists y.} \| A \|^{x,y} \otimes \| B \|^{y,z}$};
\node (apb) at (4em,-1.5em) {$\| C \|^{x,z}$};
\node (bc) at (-3.0em,5em) {$ \forall z. \| B \|^{y,z} \multimap \| C \|^{x,z}$};
\node (a) at (0em,6.5em) {$\| A \|^{x,y}$};
\node (b) at (9em,6.5em) {$\| B \|^{y,z}$};
\node (ac) at (10.6em,5em) {$\forall x. \| A \|^{x,y} \multimap \| C \|^{x,z}$};
\draw (c) -- (bc);
\draw (c) -- (ac);
\draw (ac) -- (bc);
\node at (22em,3.8em) {$A$};
\node at (22em,2.4em) {$C\mathbin{/} B$};
\node at (26em,3.8em) {$B$};
\node at (26em,2.4em) {$A\mathbin{\backslash} C$};
\node at (24em,-1em) {$A\mathbin{\bullet}B$};
\node at (24em,-2.4em) {$C$};
\node at (20em,1.6em) {$x$};
\node at (24em,1.6em) {$y$};
\node at (28em,1.6em) {$z$};
\draw (20em,1em) -- (20em,1.2em);
\draw (24em,1em) -- (24em,1.2em);
\draw (28em,1em) -- (28em,1.2em);
\draw (20em,0em) -- (20em,1em) -- (24em,1em) -- (24em,0em) -- cycle;
\draw [fill=lightgray] (24em,0em) -- (24em,1em) -- (28em,1em) -- (28em,0em) -- cycle; 	
\end{tikzpicture}
\end{center}
\caption{Figure~\ref{fig:visres} with the corresponding translations in first-order logic}
\label{fig:visresm}
\end{figure}
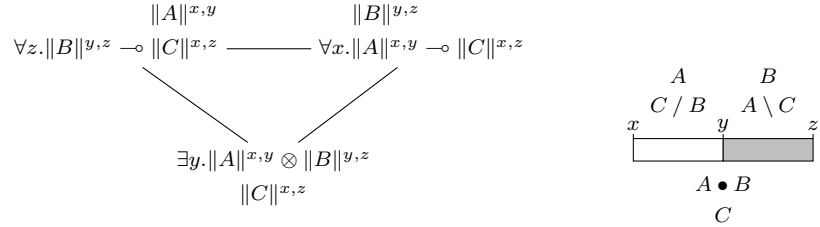

Figure~\ref{fig:visresm} shows how this translation forms a residuated triple\footnote{To show this in full detail would require us to do the simple but tedious job of proving that this definition satisfies the monotonicity and Application/Co-Application principles of Table~\ref{tab:dosen}.}. Note how combining (the translations of) $A$ and $B$ to $A\bullet B$, $A$ and $A\mathbin{\backslash} C$ to $B$, and $C\mathbin{/}{B}$ and $B$ to $C$ all correspond to the concatenation of an $x,y$ segment to an $y,z$ segment to form $x,z$ segment.

%
%
%

\subsection{Proof Nets}
\label{sec:pn}

Multiplicative linear logic has an attractive, graph-based representation of proofs called proof nets. It is relatively simple to add the first-order quantifiers to proof nets \parencite{quant,empires}.

The choice for intuitionism is justified by our interest in natural language semantics: the Curry-Howard isomorphism between proofs in multiplicative intuitionistic linear logic and linear lambda terms gives us a simple and principled way of defining the syntax-semantics interface, thereby connecting our grammatical analyses to formal linguistic semantics in the tradition of \textcite{montague}.

Proof nets can be defined in two different ways.
\begin{enumerate}
\item	
 We can define them \emph{inductively} as instructions of how to build proof nets from simpler ones.
\item We can define proof nets 
 as instances of a more general class of objects called \emph{proof structures}. 
\end{enumerate}
 
 Even though the inductive definition of proof nets is useful for proving all proof nets have certain properties, it is not immediately obvious how to determine whether something is or is not a proof net, since its inductive structure is not immediately visible (unlike, say, for sequent proofs). But to distinguish proof nets we only care about the final graph structure, the inductive structure is irrelevant\footnote{Another way of thinking about this is that different ways of producing the same proof net correspond to rule permutations in the sequent calculus.}.
 
 The second way of producing proof nets starts from proof structures. Given a sequent, there is a very direct procedure to enumerate its proof structures. Not all these proof structures will be proof nets (that is, correspond to the inductive definition of proof nets, or, equivalently, to provable sequents). A correctness condition allows us to distinguish the proof nets from other structures.

  
 Proof structures are built from the links shown in Table~\ref{tab:links}. The formulas drawn above the links are called its premisses and the formulas drawn below it are called its conclusions. Each connective is assigned two links: one where it occurs as a premiss (the left link, corresponding to the left rule for the connective in the sequent calculus) and one where it occurs as a conclusion (corresponding to the right rule in the sequent calculus).

We call the formula occurrence containing the main connective of a link its \emph{main} formula and all other formula occurrences its \emph{active} formulas.

The logical links are divided into four groups:
\begin{enumerate}
\item the \emph{tensor links} are the binary rules drawn with solid lines (the negative link for `$\multimap$' and the positive link for `$\otimes$'),
\item the \emph{par links} are the binary rules drawn with dashed lines (the negative link for `$\otimes$' and the positive link for `$\multimap$'; \emph{par} is the name for the multiplicative, classical disjunction of linear logic, `$\parr'$),
\item the \emph{existential links} are the unary rules drawn with solid lines (the negative link for `$\forall$' and the positive link for `$\exists$'),
\item  	the \emph{universal links} are the unary rules drawn with dashed lines and labeled with the corresponding eigenvariable (the negative link for `$\exists$' and the positive link for `$\forall$').
\end{enumerate}

\begin{table}
\begin{center}
\begin{tikzpicture}
\node (forallnc) at (40em,40em) {$A[x:=t]$};
\node (forallnp) [above=2em of forallnc] {$\forall x. A$};
\draw (forallnc) -- (forallnp);
\node (forallpc) [right=7em of forallnc] {$\forall x. A$};
\node (forallpp) [above=2em of forallpc] {$A$};
\draw[<-,semithick,dotted] (forallpc) -- (forallpp) node [midway] {$x\
  \ \ \ $};
%
\node (existsnc) [right=7em of forallpc] {$A$};
\node (existsnp) [above=2em of existsnc] {$\exists x. A$};
\draw[<-,semithick,dotted] (existsnp) -- (existsnc) node [midway] {$x\
  \ \ \ $};
\node (pmid) [above=7em of existsnc] {$\quad$};
\node (otimesnc) [above=2.5em of pmid] {$A\otimes B$};
\node (tmponl) [left=0.66em of otimesnc] {};
\node (aotimesnc) [left=0.66em of pmid] {$A$};
\node (tmponr) [right=0.66em of otimesnc] {};
\node (botimesnc) [right=0.66em of pmid] {$B$};
\begin{scope}
\begin{pgfinterruptboundingbox}
\path [clip] (otimesnc.center) circle (2.5ex) [reverseclip];
\end{pgfinterruptboundingbox}
\draw [semithick,dotted] (otimesnc.center) -- (botimesnc);
\draw [semithick,dotted] (otimesnc.center) -- (aotimesnc);
\end{scope}
\begin{scope}
\path [clip] (aotimesnc) -- (otimesnc.center) -- (botimesnc);
\draw (otimesnc.center) circle (2.5ex);
\end{scope}
\node (otimespc) [right=7em of pmid] {$A\otimes B$};
\node (tmpopl) [left=0.66em of otimespc] {};
\node (aotimespc) [above=2.5em of tmpopl] {$A$};
\node (tmpopr) [right=0.66em of otimespc] {};
\node (botimespc) [above=2.5em of tmpopr] {$B$};
\draw (otimespc) -- (aotimespc);
\draw (otimespc) -- (botimespc);
\node (existspc) [below=7em of otimespc] {$\exists x. A$};
\node (existspp) [above=2em of existspc] {$A[x:=t]$};
\draw  (existspc) -- (existspp);
\node (lollinc) [above=7em of forallnc] {$B$};
\node (spc) [above=7em of forallnc] {$\quad$};
\node (tmplnl) [left=0.66em of spc] {};
\node (alollin) [above=2.5em of tmplnl] {$A$};
\draw (lollinc) -- (alollin);
\node (tmplnr) [right=0.66em of spc] {};
\node (blollin) [above=2.5em of tmplnr] {$A\multimap B$};
\draw (lollinc) -- (blollin);
\node (cpos) [above=7em of forallpc] {$\quad$};
\node (lollipc) [right=0.66em of cpos] {$\phantom{(}$};
\node (aib) [right=0.5em of cpos] {$A\multimap B$};
\node (alollip) [left=0.66em of cpos] {$A$};
\node (blollip) [above=2.5em of cpos] {$B$};
\begin{scope}
\begin{pgfinterruptboundingbox}
\path [clip] (lollipc) circle (2.5ex) [reverseclip];
\end{pgfinterruptboundingbox}
\draw [semithick,dotted] (lollipc) -- (blollip);
\draw [semithick,dotted] (lollipc) -- (alollip);
\end{scope}
\begin{scope}
 \path [clip] (alollip) -- (lollipc.center) -- (blollip);
\draw (lollipc.center) circle (2.5ex);
\end{scope}
\end{tikzpicture}
\end{center}

\caption{Logical links for MILL1 proof structures}
\label{tab:twosidedlinks}
\label{tab:links}
\end{table}

\begin{definition} A \emph{proof structure} is a tuple $\mathcal{S} = \langle F, L\rangle$ where $F$ is a set of formula occurrences and $L$ is a set of the links connecting these formula occurrences such that each local subgraph is an instantiation one of the links in Table~\ref{tab:links} (for some $A$, $B$, $x$, $t$), and such that
\begin{itemize}
\item each formula is at most once the premiss of a link,
\item each formula is at most once the conclusion of a link.
\end{itemize}

Finally, the quantifiers links and eigenvariables have the following additional conditions.
\begin{itemize}
\item each quantifier link uses a distinct bound variable,
\item all conclusions and hypotheses of $\mathcal{S}$ are closed,
\item all eigenvariables of links in $\mathcal{S}$ are used strictly, meaning that we cannot substitute a constant $c_x$ for any set of occurrences of an eigenvariable $x$ and obtain a proof structure with the same conclusions and hypotheses.
\end{itemize}

The formulas which are not the premisses of any link in a proof structure with hypotheses are the \emph{conclusions} of the structure. The formulas which are not the conclusions of any link are the \emph{hypotheses} of the structure.

Formulas which are both the premiss and the conclusion of a link in a proof structure are its \emph{internal} formulas. All other formulas (that is, formulas which are either hypotheses or conclusions of the proof structure) are its \emph{external} formulas.
\defend
\end{definition}

This definition essentially follows \textcite{quant}, incorporating the notion of strictly used eigenvariables from \textcite{empires} and the proof structures with hypotheses of \textcite{reductions}. The requirement that eigenvariables are used strictly avoids the case where, for example, a subproof $\forall x.a(x)\vdash \exists y.a(y)$ instantiates $x$ and $y$ to the eigenvariable $z$ of a universal link elsewhere in the proof. Given that, by definition, we can replace such occurrences by a new constant $c_z$ this is a minor technicality to facilitate the verification of the correctness of the universal links in a proof net.

\begin{figure}
\begin{center}
\begin{tikzpicture}
\node (lollinc) {$a\otimes \forall x. b(x)$};
\node (tmplnl) [left=0.0em of lollinc] {};
\node (tmplnr) [right=0.0em of lollinc] {};
\node (blollin) [above=2.5em of tmplnr] {$\forall x. b(x)$};
\draw (lollinc) -- (blollin);
%
%
\node (blollip) [above=2em of blollin] {$b(x)$};
\draw[<-,semithick,dotted] (blollin) -- (blollip) node [midway] {$x\
  \ \ \ $};
\node (alollin) [above=6.3em of tmplnl] {$a$};
\draw (lollinc) -- (alollin);
\node (existspp) [above=9.5em of lollinc]{$a\otimes b(x)$};
\begin{scope}
\begin{pgfinterruptboundingbox}
\clip (existspp.center) circle (2.5ex) [invclip];
\end{pgfinterruptboundingbox}
\draw [semithick,dotted] (existspp.center) -- (blollip);
\draw [semithick,dotted] (existspp.center) -- (alollin);
\end{scope}
\begin{scope}
\path [clip] (alollin) -- (existspp.center) -- (blollip);
\draw (existspp.center) circle (2.5ex);
\end{scope}
\node (hyp) [above=2.0em of existspp] {$\forall y. [a\otimes b(y)]$};
\draw (hyp) -- (existspp);
\node (tmpb) [left=12em of lollinc] {};
\node (lollincb) [below=1em of tmpb] {$a\otimes \forall x. b(x)$};
\node (tmplnl) [left=0.0em of lollincb] {};
\node (tmplnr) [right=0.0em of lollincb] {};
\node (blollin) [above=2.5em of tmplnr] {$\forall x. b(x)$};
\draw (lollincb) -- (blollin);
%
%
\node (bbis) [above=2.5em of blollin] {$b(x)$};
\node (blollip) [above=2em of bbis] {$b(Y)$};
\draw[<-,semithick,dotted] (blollin) -- (bbis) node [midway] {$x\
  \ \ \ $};
\node (alollin) [above=10.6em of tmplnl] {$a$};
\node (abis) [below=6.8em of alollin] {$a$};
\draw (lollincb) -- (abis);
\node (existspp) [above=13.0em of lollincb]{$a\otimes b(Y)$};
\begin{scope}
\begin{pgfinterruptboundingbox}
\clip (existspp.center) circle (2.5ex) [invclip];
\end{pgfinterruptboundingbox}
\draw [semithick,dotted] (existspp.center) -- (blollip);
\draw [semithick,dotted] (existspp.center) -- (alollin);
\end{scope}
\begin{scope}
\path [clip] (alollin) -- (existspp.center) -- (blollip);
\draw (existspp.center) circle (2.5ex);
\end{scope}
\node (hyp) [above=2.0em of existspp] {$\forall y. [a\otimes b(y)]$};
\draw (hyp) -- (existspp);

%
%

\end{tikzpicture}
\end{center}
\caption{Two proof structures for the sequent $\forall y [a \otimes b(y)] \vdash a\otimes
  \forall x.b(x)$.}
\label{fig:twoslpn}
\end{figure}
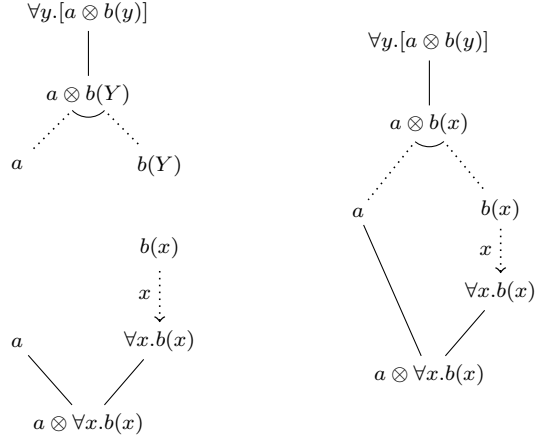

Figure~\ref{fig:twoslpn} shows, on the left hand side, the formula unfolding for the underivable sequent $\forall y [a \otimes b(y)] \vdash a\otimes
  \forall x.b(x)$. We want derivable sequents $A_1,\ldots,A_n\vdash C$ to correspond to proof structures (and proof \emph{nets}) with exactly the $A_i$ as hypotheses and $C$ as a conclusion. The proof structure on the left hand side of Figure~\ref{fig:twoslpn} has $a$ and $b(Y)$ as additional conclusions and $a$ and $b(x)$ as additional hypotheses. By identifying these formulas (and substituting $x$ for $Y$) we obtain the proof structure shown on the right hand side of Figure~\ref{fig:twoslpn}. In the current case, this is the unique identification of atomic formulas producing a proof structure such that the only hypothesis is $\forall y [a\otimes b(y)]$ and the only conclusion is $a\otimes \forall x. b(x)$. In the general case, there can be many ways of identifying atomic formulas and this is the central problem for proof search using proof nets.
    
Underivability in the sequent calculus follows from the fact that there is no proof where the $\forall$ right rule is performed below the $\forall$ left rule (the intuitionistic version of this sequent $\forall y [a \wedge b(y)] \vdash a\wedge
  \forall x.b(x)$ \emph{is} derivable, but it requires us to use the antecedent formula $\forall y [a \wedge b(y)]$ twice, which produces the correct order between the $\forall$ left and right rules). We will see below why the proof structure on the right of Figure~\ref{fig:twoslpn} is not a proof net.

\begin{definition} Given a proof structure $\mathcal{P}$ a \emph{component} is a maximal, connected substructure containing only tensor and existential links.
\end{definition}

We obtain the components of a proof structure by first removing the par and universal links, then taking each (maximal) connected substructure. Components can be single formulas. 
The components of the proof structure on the right of Figure~\ref{fig:twoslpn} correspond to the induced substructures of 
$\{ \forall y. [a\otimes b(y)], a \otimes b(x) \}$,
     $\{ a, \forall x. b(x), a\otimes \forall x. b(x)\}$, and $\{ b(x) \}$. For the first and last of these structures, the occurrences of $x$ (all of them free) will be replaced by $c_x$. The second substructure contains the universal link for $x$ (and only bound occurrences of $x$) and its formulas will therefore be unchanged. 
     The corresponding  sequents are given in Equations~\ref{eq:cnp1} to~\ref{eq:cnp99}.
\begin{align}
\label{eq:cnp1} \forall
    y. [a\otimes b(y)] &\vdash a \otimes b(c_x) \\
a, \forall x. b(x)   &\vdash a\otimes \forall x. b(x)\\
b(c_x) &\vdash b(c_x) \label{eq:cnp99}	
\end{align}

The reader can verify that all of these are derivable (though we cannot combine these three proofs into a single proof of the required endsequent). Before we turn to the correctness condition, we need another auxiliary notion from \textcite{empires}.

\begin{definition} Given a proof structure $\mathcal{P}$ and the eigenvariable $x$ of a link $l$ in $P$, the \emph{existential frontier} of $x$ in $\mathcal{P}$ is the set of formula occurrences $A_1,\ldots,A_n$ such that each $A_i$ is the main formula of an existential link $l_i$ where $x$ occurs free in the active formula of $l_i$ but not in its main formula $A_i$.
\end{definition}

In Figure~\ref{fig:twoslpn}, the formula $\forall
    y. [a\otimes b(y)]$ is the only formula in the existential frontier of $x$.
    
To decide whether a proof structure is a proof net in linear logic, we need a correctness condition on the proof structure. Given that the two universal links correspond to sequent calculus rules with side conditions on the use of their eigenvariable, it should come as no surprise that we need to keep track of free occurrences of eigenvariables for deciding correctness. Typical correctness conditions involve graph switchings and graph contractions. \textcite{quant}, and \textcite{empires} extend the switching condition of \textcite{multiplicatives} for first-order linear logic. Here we will extend the contraction condition of \textcite{reductions} to the first-order case.

\begin{definition} An abstract proof structure $\mathcal{A}=\langle V,L\rangle$ is obtained from a proof structure $\mathcal{P}=\langle F,L\rangle$ by replacing each formula $A \in F$ by the set of eigenvariables freely occurring in $A$, plus the eigenvariable $x$ in case $A$ is on the existential frontier of a universal link of $\mathcal P$.
\end{definition}

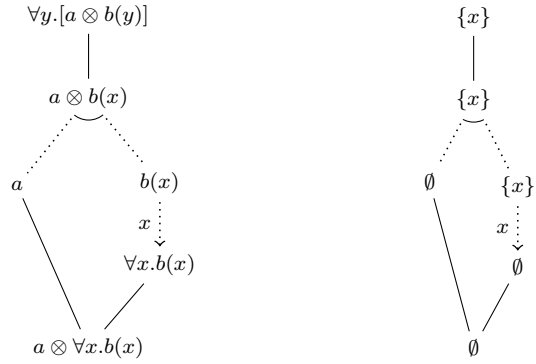
\begin{figure}
\begin{center}
\begin{tikzpicture}
\node (lollinc) {$a\otimes \forall x. b(x)$};
\node (tmplnl) [left=0.0em of lollinc] {};
\node (tmplnr) [right=0.0em of lollinc] {};
\node (blollin) [above=2.5em of tmplnr] {$\forall x. b(x)$};
\draw (lollinc) -- (blollin);
%
%
\node (blollip) [above=2em of blollin] {$b(x)$};
\draw[<-,semithick,dotted] (blollin) -- (blollip) node [midway] {$x\
  \ \ \ $};
\node (alollin) [above=6.3em of tmplnl] {$a$};
\draw (lollinc) -- (alollin);
\node (existspp) [above=9.5em of lollinc]{$a\otimes b(x)$};
\begin{scope}
\begin{pgfinterruptboundingbox}
\clip (existspp.center) circle (2.5ex) [invclip];
\end{pgfinterruptboundingbox}
\draw [semithick,dotted] (existspp.center) -- (blollip);
\draw [semithick,dotted] (existspp.center) -- (alollin);
\end{scope}
\begin{scope}
\path [clip] (alollin) -- (existspp.center) -- (blollip);
\draw (existspp.center) circle (2.5ex);
\end{scope}
\node (hyp) [above=2.0em of existspp] {$\forall y. [a\otimes b(y)]$};
\draw (hyp) -- (existspp);
\node (tmp) [right = 12em of lollinc] {};
\node (lollinc) [right=0em of tmp] {$\quad\emptyset\quad$};
\node (tmplnl) [left=0.0em of lollinc] {};
\node (tmplnr) [right=0.0em of lollinc] {};
\node (blollin) [above=2.5em of tmplnr] {$\quad\emptyset\quad$};
\draw (lollinc) -- (blollin);
%
%
\node (blollip) [above=2em of blollin] {$\{ x \}$};
\draw[<-,semithick,dotted] (blollin) -- (blollip) node [midway] {$x\
  \ \ \ $};
\node (alollin) [above=6.3em of tmplnl] {$\quad\emptyset\quad$};
\draw (lollinc) -- (alollin);
\node (existspp) [above=9.5em of lollinc]{$\{ x \}$};
\begin{scope}
\begin{pgfinterruptboundingbox}
\clip (existspp.center) circle (2.5ex) [invclip];
\end{pgfinterruptboundingbox}
\draw [semithick,dotted] (existspp.center) -- (blollip);
\draw [semithick,dotted] (existspp.center) -- (alollin);
\end{scope}
\begin{scope}
\path [clip] (alollin) -- (existspp.center) -- (blollip);
\draw (existspp.center) circle (2.5ex);
\end{scope}
\node (hyp) [above=2.0em of existspp] {$\{ x \}$};
\draw (hyp) -- (existspp);

\end{tikzpicture}
\end{center}
\caption{Proof structure (left) and abstract proof structure (right) for the sequent $\forall y [a \otimes b(y)] \vdash a\otimes
  \forall x.b(x)$.}
\label{fig:twosapn}
\end{figure}

Figure~\ref{fig:twosapn} shows the proof structure and corresponding abstract proof structure of the proof structure we've seen before on the right of Figure~\ref{fig:twoslpn}.
We have simply erased the formula information and kept only the information of the free variables at each node. The top node and only hypothesis of the structure, which corresponds to a closed formula (the formula $\forall
    y. [a\otimes b(y)]$), is on the existential frontier of $x$ (there is an occurrence of $x$ in the active formula of the link) and therefore has the singleton set $\{ x \}$ assigned to it.

Table~\ref{fig:controne} shows the contractions for first-order linear logic. 
Each contraction is an edge contraction on the abstract proof structure, deleting an edge or a joined pair of edges, and identifying the two incident vertices $v_i$ and $v_j$. The resulting vertex is incident both to all nodes incident to $v_i$ (except $v_j$) and to all nodes incident to $v_j$ (except $v_i$). The eigenvariables assigned to the resulting vertex are the set union of the eigenvariables assigned to $v_i$ and $v_j$. For the universal contraction $\textit{u}$ the eigenvariable corresponding to the eigenvariable $x$ of the link is removed.  
The contraction $\textit{p}$ verifies that the two premisses of a single par link can be joined in a single point. The contraction $\textit{u}$ verifies that all free occurrences of the eigenvariable of a universal link (and its existential border) can be found at the vertex corresponding to the premiss of the link. The contraction $\textit{c}$ contracts a component.

All contractions remove one edge (or, in the case of the par contraction $\textit{p}$, a linked pair of edges) and keep all other edges the same, reducing the length of the paths which passed through the contracted edge by one.
Contractions can produce self-loops and multiple edges between two nodes, but can never remove self-loops. 

\begin{table}
\begin{center}
\begin{tikzpicture}
\node (x) at (0em,0em) {$v_i$};
\node (y) at (0em,4em) {$v_j$};
\draw [semithick,dotted] plot [smooth, tension=1] coordinates {(-0.4em,0.5em) (-1em,2em) (-0.4em,3.5em)};
\draw [semithick,dotted] plot [smooth, tension=1] coordinates {(0.4em,0.5em) (1em,2em) (0.4em,3.5em)};
\draw plot [smooth,tension=1] coordinates {(-0.4em,0.5em) (0em,0.7em) (0.4em,0.5em)};
%
\node (x2) at (4em,2em) {$v_i$};
\node (a1) at (2.5em,2em) {$\Rightarrow_{\textit{p}}$};
\node (x3) at (10em,0em) {$v_i$};
\node (y3) at (10em,4em) {$v_j$};
\draw[<-,semithick,dotted] (x3) -- (y3);
\node (xv) at (9.45em,2.15em) {$x$};
\node (x4) at (14em,2em) {$v_i$};
\node (a2) at (12em,2em) {$\Rightarrow_{\textit{u}}$};
\node (x3) at (20em,0em) {$v_i$};
\node (y3) at (20em,4em) {$v_j$};
\draw (x3) -- (y3);
\node (x4) at (24em,2em) {$v_i$};
\node (a3) at (22em,2em) {$\Rightarrow_{\textit{c}}$};
\end{tikzpicture}
\end{center}
\caption{Contractions for first-order linear logic. Conditions: $v_i \neq v_j$
  and, for the $u$ contraction, all occurrences of $x$ are at $v_j$.}
\label{fig:controne}
\end{table}

\begin{definition} A proof structure is a \emph{proof net} iff its abstract proof structure contracts to a single vertex using the contractions of Table~\ref{fig:controne}.
\end{definition}

\begin{figure}
\begin{center}
\begin{tikzpicture}
\node (lollinc) {$\quad\emptyset\quad$};
\node (tmplnl) [left=0.0em of lollinc] {};
\node (tmplnr) [right=0.0em of lollinc] {};
\node (blollin) [above=2.5em of tmplnr] {$\quad\emptyset\quad$};
\draw (lollinc) -- (blollin);
%
%
\node (blollip) [above=2em of blollin] {$\{ x \}$};
\draw[<-,semithick,dotted] (blollin) -- (blollip) node [midway] {$x\
  \ \ \ $};
\node (alollin) [above=6.3em of tmplnl] {$\quad\emptyset\quad$};
\draw (lollinc) -- (alollin);
\node (existspp) [above=9.5em of lollinc]{$\{ x \}$};
\begin{scope}
\begin{pgfinterruptboundingbox}
\clip (existspp.center) circle (2.5ex) [invclip];
\end{pgfinterruptboundingbox}
\draw [semithick,dotted] (existspp.center) -- (blollip);
\draw [semithick,dotted] (existspp.center) -- (alollin);
\end{scope}
\begin{scope}
\path [clip] (alollin) -- (existspp.center) -- (blollip);
\draw (existspp.center) circle (2.5ex);
\end{scope}
\node (hyp) [above=2.0em of existspp] {$\{ x \}$};
\draw (hyp) -- (existspp);
\node (ar1) [right=1.5em of blollip] {$\rightarrow$};
\node (tmp) [right=7em of lollinc] {};
\node (lollinc) [right=0em of tmp] {$\quad\ \quad$};
\node (tmplnl) [left=0.0em of lollinc] {};
\node (tmplnr) [right=0.0em of lollinc] {};
\node (blollin) [above=2.5em of tmplnr] {$\quad\emptyset\quad$};
%
%
\node (blollip) [above=2em of blollin] {$\{ x \}$};
\draw[<-,semithick,dotted] (blollin) -- (blollip) node [midway] {$x\
  \ \ \ $};
\node (alollin) [above=6.3em of tmplnl] {$\quad\emptyset\quad$};
\draw (blollin) -- (alollin);
\node (existspp) [above=9.5em of lollinc]{$\{ x \}$};
\begin{scope}
\begin{pgfinterruptboundingbox}
\clip (existspp.center) circle (2.5ex) [invclip];
\end{pgfinterruptboundingbox}
\draw [semithick,dotted] (existspp.center) -- (blollip);
\draw [semithick,dotted] (existspp.center) -- (alollin);
\end{scope}
\begin{scope}
\path [clip] (alollin) -- (existspp.center) -- (blollip);
\draw (existspp.center) circle (2.5ex);
\end{scope}
\node (hyp) [above=2.0em of existspp] {$\{ x \}$};
\draw (hyp) -- (existspp);
\node (ar2) [right=1.5em of blollip] {$\rightarrow$};
\node (tmp) [right=7em of lollinc] {};
\node (lollinc) [right=0em of tmp] {$\quad\ \quad$};
\node (tmplnl) [left=0.0em of lollinc] {};
\node (tmplnr) [right=0.0em of lollinc] {};
\node (blollin) [above=2.5em of tmplnr] {$\quad\ \quad$};
%
%

%
\node (alollin) [above=6.3em of tmplnl] {$\emptyset$};
\node (blollip) [right=2.5em of alollin] {$\{ x \}$};
%
\node (existspp) [above=9.5em of lollinc]{$\{ x \}$};
\draw[<-,semithick,dotted] (alollin) -- (blollip) node [midway] {$x_{\rule{0pt}{1.5em}}$};
\begin{scope}
\begin{pgfinterruptboundingbox}
\clip (existspp.center) circle (2.5ex) [invclip];
\end{pgfinterruptboundingbox}
\draw [semithick,dotted] (existspp.center) -- (blollip);
\draw [semithick,dotted] (existspp.center) -- (alollin);
\end{scope}
\begin{scope}
\path [clip] (alollin) -- (existspp.center) -- (blollip);
\draw (existspp.center) circle (2.5ex);
\end{scope}
\node (hyp) [above=2.0em of existspp] {$\{ x \}$};
\draw (hyp) -- (existspp);
\node (ar3) [right=1.5em of blollip] {$\rightarrow$};
\node (tmp) [right=7em of lollinc] {};
\node (lollinc) [right=0em of tmp] {$\quad\ \quad$};
\node (tmplnl) [left=0.0em of lollinc] {};
\node (tmplnr) [right=0.0em of lollinc] {};
\node (blollin) [above=2.5em of tmplnr] {$\quad\ \quad$};
%
%

%
\node (alollin) [above=6.3em of tmplnl] {$\emptyset$};
\node (blollip) [right=2.5em of alollin] {$\{ x \}$};
%
\node (existspp) [above=9.5em of lollinc]{$\{ x \}$};
\draw[<-,semithick,dotted] (alollin) -- (blollip) node [midway] {$x_{\rule{0pt}{1.5em}}$};
\begin{scope}
\begin{pgfinterruptboundingbox}
\clip (existspp.center) circle (2.5ex) [invclip];
\end{pgfinterruptboundingbox}
\draw [semithick,dotted] (existspp.center) -- (blollip);
\draw [semithick,dotted] (existspp.center) -- (alollin);
\end{scope}
\begin{scope}
\path [clip] (alollin) -- (existspp.center) -- (blollip);
\draw (existspp.center) circle (2.5ex);
\end{scope}
\end{tikzpicture}
\end{center}
\caption{Failed contraction sequence for the abstract proof structure on the right of Figure~\ref{fig:twoslpn}}
\label{fig:failed}
\end{figure}
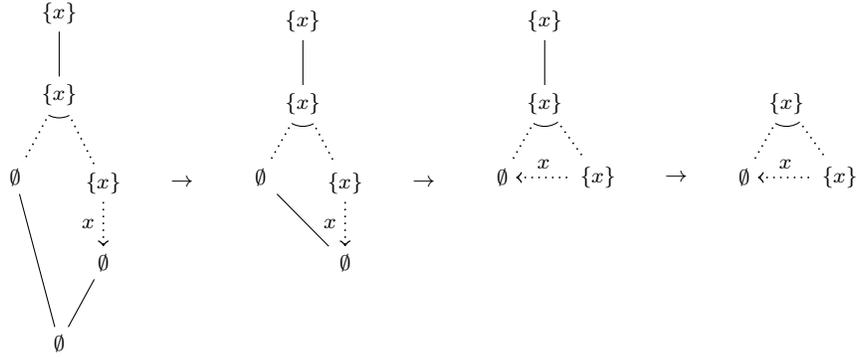

The contraction system as presented is not confluent. For the critical cases, when a pair of vertices $v_1$ and $v_2$ is connected by two or more links of different types (par, universal or component), we can contract any of these multiple links connective $v_1$ and $v_2$ and produce a self-loop for all others. An easy solution to ensure confluence is to treat all self-loops as equivalent\footnote{A more elegant solution for ensuring confluence would replace the right-hand side of the $\mathit{p}$ and $\mathit{u}$ contractions by the left-hand side of the $\mathit{c}$ contraction.}. Figure~\ref{fig:failed} shows how the abstract proof structure of Figure~\ref{fig:twoslpn} fails to contract to a single vertex. The final structure shown on the right of the figure cannot be further contracted: the par (p) contraction requires the two edges of the par link to end in the same vertex, whereas the universal (u) contraction requires all occurrences of $x$ to be at the vertex from which the $x$ edge is leaving.

\begin{lemma} $\Gamma\vdash C$ is derivable if and only if there is a proof net of $\Gamma\vdash C$	
\end{lemma}

See \textcite{empires} for a proof, which adapts trivially to the current context.

\section{Residuation and Partial Orders}

So far, we have discussed proof-theoretic properties of first-order linear logic while only hinting at its applications as a formalism for natural language processing. In this section, I will suggest some principles for writing grammars using first-order linear logic, essentially in the form of constraints on the formulas. These constraints apply only to constants and variables used as string positions and not to other applications of first-order variables (such as grammatical case, island constraints and scoping constraints). The principles presented here should not be taken in a dogmatic way.  It may turn out that a larger class of grammars has significant applications or better mathematical properties. The goal is merely to provide some terra firma for exploring both linguistic applications and mathematical properties. Indeed, some known classes of type-logical grammars are outside the residuated fragment investigated in this paper \parencite{oehrle}, even though it is possible to follow \textcite{kl20tls} and \emph{combine} residuated connectives with non-residuated ones in the more general framework proposed here.

The main property we want our formulas to preserve is that we can always uniquely define a linear order on the string segments (pairs of position variables) used in the formulas of first-order linear logic. This is already somewhat of a shift with respect to standard first-order linear logic: an atomic formula $p(x_0,x_1,x_2,x_3)$ represents to string segments $x_0,x_1$ and $x_2,x_3$ without any claims about the relative order of these two segments. This gives us the freedom to build these two strings independently and let other lexical items in the grammar decide in which relative order these two segments will ultimately appear in the derived string. Adding the linear order requirement requires us to add an explicit relation between these two segments (either $x_1 \leq x_2$, for the linear order $x_0,x_1,x_2,x_3$, or $x_3 \leq x_0$ for the linear order $x_2,x_3,x_0,x_1$). 

It is possible to define residuated connectives for string segments which are not linearly ordered. However, we would then be limited by the fact that any connective which linearises such segments (by ordering some of the previously unordered segments) would not be residuated. For example, suppose we want to define a connective combining two unordered string segments $x_0,x_1$ and $x_2,x_3$ by concatenating  them (or `wrapping' them around) a segment $x_1,x_2$ producing the complex segment $x_0,x_1$. This would entail the linear order to be $x_0,x_1,x_2,x_3$, and therefore the two segments $x_0,x_1$ and $x_2,x_3$ assigned to one of the residuals must be linearly ordered as well, simply because the alternative order $x_2,x_3,x_0,x_1$ has become incompatible with the linear order after concatenation. A restriction to residuated connective therefore sacrifices some flexibility for writing grammars in first-order linear logic. We will return briefly to this point in the discussion of Section~\ref{sec:discussion}.

\subsection{Residuation for the Lambek Calculus Revisited}

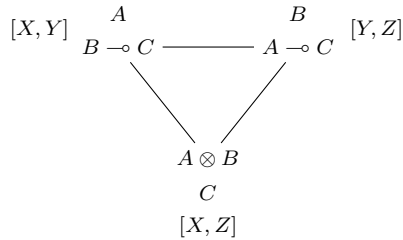
\begin{figure}
\begin{center}
\begin{tikzpicture}
\node (c) at (4em,0em) {$A\otimes B$};
\node (apb) at (4em,-1.5em) {$C$};
\node (bc) at (0em,5em) {$B\multimap C$};
\node (a) at (0em,6.5em) {$A$};
\node (b) at (8em,6.5em) {$B$};
\node (ac) at (8em,5em) {$A\multimap C$};
\draw (c) -- (bc);
\draw (c) -- (ac);
\draw (ac) -- (bc); 	
\node (apos) at (-3.5em,5.75em) {$\poslin{X,Y}$};
\node (bpos) at (11.5em,5.75em) {$\poslin{Y,Z}$};
\node (cpos) at (4em,-3em) {$\poslin{X,Z}$};
\end{tikzpicture}
\end{center}
\caption{Lambek calculus residuation translated into first-order linear logic.}
\label{fig:reslam}
\end{figure}

We have already looked at the Lambek calculus connectives and their translation into linear logic from the point of view of residuation. 
 Figure~\ref{fig:reslam} presents a simplified version of Figure~\ref{fig:visresm}. It focuses only on the position variables, which have been placed at the appropriate points in the triangle.

Each variable occurs on exactly two of the tree points of the triangle. The place where a variable is absent determines the quantifier: `$\exists$' for `$\otimes$' (that is, the bottom node), and `$\forall$' for the two `$\multimap$' nodes (the two top nodes). Downwards movement --- from $A$ and $B$ to $A\otimes B$, from $A$ and $A\multimap C$ to $C$, and from $B\multimap C$ and $B$ to $C$ --- corresponds to concatenation: we combine a first string with left position $X$ and right position $Y$ with a second string with left position $Y$ and right position $Z$ to form a new string starting at the left position $X$ of the first and ending at the right position $Z$ of the second.

A variable shared between the bottom position and one of the top positions of the figure must appear in both of these in either a left position or a right position (as, respectively, variables $X$ and $Z$ in Figure~\ref{fig:reslam}).
A variable shared among the two top positions must appear in a right position in one and a left position in the other. Variable $Y$ in the figure is in this case.

Seen from the point of view of string segments, the bottom element contains exactly the combination of the string segments of the left and right elements, with some of them (that is those positions occurring both left and right) concatenated.


\editout{
From $A\poslin{X,Y}$ and $C\poslin{X,Z}$ we conclude $(A\multimap C)\poslin{Y,Z}$. But why not conclude $(A\multimap C)\poslin{Z,Y}$ instead? If $X$ occurs to the left of both $Y$ and $Z$ that tells us nothing about the relative position of $Y$ and $Z$ in a linear order. However, since the combination of $A$ and $A\multimap C$ is concatenation-like, only the $Y,Z$ order makes sense. Or, the result category $C$ cannot be smaller than any of its arguments.
}

\subsection{Partial Orders}

As a general principle, we want the left-to-right order of the position variables and constants to be globally coherent. This means that we do not want $X$ to be left of $Y$ at one place and to the right of it at another (at least not unless they are equal). Formally, this means that the variables in a formula and in a proof are partially ordered. More precisely, we have only argued for antisymmetry (that is $X \leq Y$ and $Y \leq X$ entail $X = Y$). To be a partial order, we also need reflexivity ($X\leq X$) and transitivity ($X\leq Y$ and $Y \leq Z$ entail $X\leq Z$, or, in our terms: if $X$ occurs to the left of $Y$ and $Y$ occurs to the left of $Z$ then $X$ occurs to the left of $Z$).


We can add explicit partial order constraints to first-order linear logic, where a lexical entry specifies explicitly how some of its variables are ordered.  In a system with explicit partial order constraints, a sequent is derivable if it is derivable in first-order linear logic (as before) but also satisfies all lexical constraints on the partial order. We will see in the next section how this can be useful.

Instead of using partial order constraints to obtain extra expressivity, we can also see it as a way of improving efficiency. For example, when we look at a sentence like.

\begin{enumerate}
\item John gave Mary flowers.
\end{enumerate}

With formulas $np$, $((np\backslash s)/np)/np$, $np$, and $np$, we obtain the formula $np(0,1)$ for ``John'' and $\forall Z. np(Y,Z) \multimap \forall Y. np(2,Y)\multimap \forall X. np(X,1) \multimap s(X,Z)$ for ``gave'' (using the standard Lambek calculus translation). This produces the orders $0 < 1$ for ``John'' and $X\leq 1 < 2 \leq Y \leq Z$ for ``gave''. Without any partial order constraints, it would be possible to identify $np(0,1)$ with $np(Y,Z)$. With the contraint, this would fail, since unifying $Y$ with $0$ would entail $2 \leq 0$ contradicting $0 < 2$. We will give a more detailed and interesting example in Section~\ref{sec:poexe}.

\editout{
Implicit partial orders are simply a way of enforcing that variables are used consistently \emph{as position variables}. When we have assigned a partial order to the variables occurring in two immediate subformulas $A$ and $B$ of a formula $A\otimes B$ or $A\multimap B$, we want to assign a partial order to the full formula. However, the union of two partial orders is not necessarily a partial order. 
 If $X\leq Y$ in $A$ and $Y\leq X$ in $B$ this would entail $X=Y$, side-stepping the condition that each variable has two occurrences in a formula.
A more complex example would be where $X\leq Y$ and $V\leq W$ in $A$ and $Y\leq V$ and $W \leq X$ in $B$.}

The residuation principle for generalised forms of concatenation requires use to be able to uniquely reconstruct the linear order of any of the three elements in a residuated triple based on the linear order of the two others.
As we will see, for three position variables and two string segments, the Lambek calculus connectives are the only available residuated triple. But what happens when we increase the number of variables, and thereby the number of string positions?

\begin{figure}
\begin{center}
\begin{tikzpicture}
\node (c) at (4em,0em) {$\smash{\exists x_1,x_2.} \| A \|^{x_0,x_1,x_2,x_3} \otimes \| B \|^{x_1,x_2}$};
\node (apb) at (4em,-1.5em) {$\| C \|^{x_0,x_3}$};
\node (bc) at (-4.1em,5em) {$  \| B \|^{x_1,x_2} \multimap \| C \|^{x_0,x_3}$};
\node (a) at (0em,6.5em) {$\| A \|^{x_0,x_1,x_2,x_3}$};
\node (b) at (9em,6.5em) {$\| B \|^{x_1,x_2}$};
\node (ac) at (11.2em,5em) {$\forall x_0, x_3. \| A \|^{x_0,x_1,x_2,x_3} \multimap \| C \|^{x_0,x_3}$};
\draw (c) -- (bc);
\draw (c) -- (ac);
\draw (ac) -- (bc);
\node at (22em,3.8em) {$A$};
\node at (22em,2.4em) {$C\mathbin{/_{3a}} B$};
\node at (30em,3.8em) {$A$};
\node at (30em,2.4em) {$C\mathbin{/_{3a}} B$};
\node at (26em,3.8em) {$B$};
\node at (26em,2.4em) {$A\mathbin{\backslash_{3a}} C$};
\node at (26em,-1em) {$A\mathbin{\bullet_{3a}}B$};
\node at (26em,-2.4em) {$C$};
\node at (20em,1.6em) {$x_0$};
\node at (24em,1.6em) {$x_1$};
\node at (28em,1.6em) {$x_2$};
\node at (32em,1.6em) {$x_3$};
\draw (20em,1em) -- (20em,1.2em);
\draw (24em,1em) -- (24em,1.2em);
\draw (28em,1em) -- (28em,1.2em);
\draw (32em,1em) -- (32em,1.2em);
\draw (20em,0em) -- (20em,1em) -- (24em,1em) -- (24em,0em) -- cycle;
\draw [fill=lightgray] (24em,0em) -- (24em,1em) -- (28em,1em) -- (28em,0em) -- cycle;
\draw (28em,0em) -- (28em,1em) -- (32em,1em) -- (32em,0em) -- cycle; 	
\end{tikzpicture}

\begin{tikzpicture}
\node (c) at (4em,0em) {$ \| A \|^{x_0,x_1} \otimes \| B \|^{x_2,x_3}$};
\node (apb) at (4em,-1.5em) {$\| C \|^{x_0,x_1,x_2,x_3}$};
\node (bc) at (-5em,5em) {$ \forall x2,x3. \| B \|^{x_2,x_3} \multimap \| C \|^{x_0,x_1,x_2,x_3}$};
\node (a) at (0em,6.5em) {$\| A \|^{x_0,x_1}$};
\node (b) at (9em,6.5em) {$\| B \|^{x_2,x_3}$};
\node (ac) at (12em,5em) {$\forall x_0,x_1.\| A \|^{x_0,x_1} \multimap \| C \|^{x_0,x_1,x_2,x_3}$};
\draw (c) -- (bc);
\draw (c) -- (ac);
\draw (ac) -- (bc);
\node at (22em,3.8em) {$A$};
\node at (22em,2.4em) {$C\mathbin{/_{3b}} B$};
\node at (30em,3.8em) {$B$};
\node at (30em,2.4em) {$A\mathbin{\backslash_{3b}} C$};
\node at (26em,-1em) {$A\mathbin{\bullet_{3b}}B$};
\node at (26em,-2.4em) {$C$};
\node at (20em,1.6em) {$x_0$};
\node at (24em,1.6em) {$x_1$};
\node at (28em,1.6em) {$x_2$};
\node at (32em,1.6em) {$x_3$};
\draw (20em,1em) -- (20em,1.2em);
\draw (24em,1em) -- (24em,1.2em);
\draw (28em,1em) -- (28em,1.2em);
\draw (32em,1em) -- (32em,1.2em);
\draw (20em,0em) -- (20em,1em) -- (24em,1em) -- (24em,0em) -- cycle;
\draw [fill=lightgray] (28em,0em) -- (28em,1em) -- (32em,1em) -- (32em,0em) -- cycle; 	
\end{tikzpicture}

\editout{
\begin{tikzpicture}
\node (c) at (4em,0em) {$A\otimes B$};
\node (apb) at (4em,-1.5em) {$C$};
\node (bc) at (0em,5em) {$B\multimap C$};
\node (a) at (0em,6.5em) {$A$};
\node (b) at (8em,6.5em) {$B$};
\node (ac) at (8em,5em) {$A\multimap C$};
\draw (c) -- (bc);
\draw (c) -- (ac);
\draw (ac) -- (bc); 	
\node (phantom) at (-4.8em,5.75em) {\phantom{$\poslin{x_0,x_1,x_2,x_3}$}};
\node (apos) at (-3.5em,5.75em) {$\poslin{x_0,x_1}$};
\node (bpos) at (11.5em,5.75em) {$\poslin{x_2,x_3}$};
\node (cpos) at (4em,-3em) {$\poslin{x_0,x_1,x_2,x_3}$};
\node (forall) at (11.5em,7.6em) {$\forall x_0,x_1$};
\node (forallb) at (-3.5em,7.6em) {$\forall x_2,x_3$};
\end{tikzpicture}
\bigskip
\begin{tikzpicture}
\node (filler) at (4em,8em) {\ };
\node (c) at (4em,0em) {$A\otimes B$};
\node (apb) at (4em,-1.5em) {$C$};
\node (bc) at (0em,5em) {$B\multimap C$};
\node (a) at (0em,6.5em) {$A$};
\node (b) at (8em,6.5em) {$B$};
\node (ac) at (8em,5em) {$A\multimap C$};
\draw (c) -- (bc);
\draw (c) -- (ac);
\draw (ac) -- (bc); 	
\node (apos) at (-4.8em,5.75em) {$\poslin{x_0,x_1,x_2,x_3}$};
\node (bpos) at (11.5em,5.75em) {$\poslin{x_1,x_2}$};
\node (cpos) at (4em,-3em) {$\poslin{x_0,x_3}$};
\node (forall) at (11.5em,7.6em) {$\forall x_0,x_3.$};
\node (exists) at (-1em,-3.1em) {$\exists x_1,x_2$};
\end{tikzpicture}}
\end{center}
\caption{Two families of connectives with three segments and four position variables}
\label{fig:fourpos}
\end{figure}
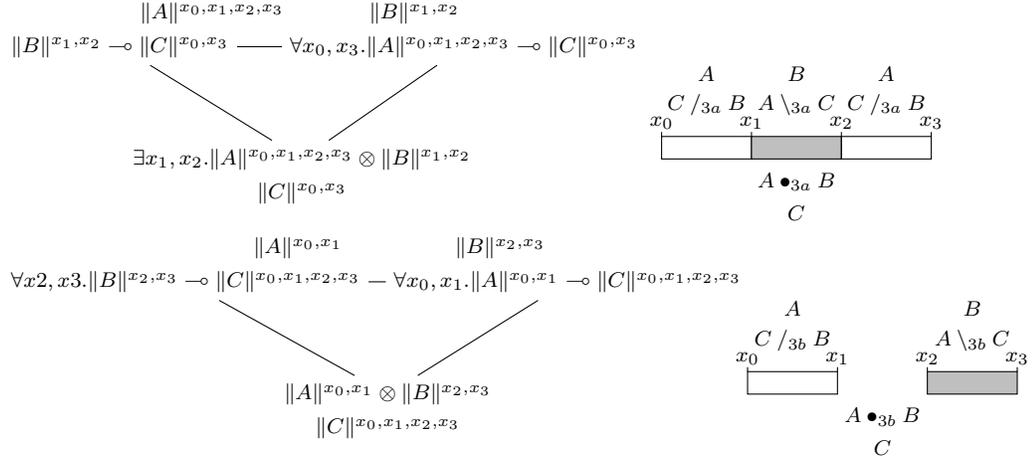

Figure~\ref{fig:fourpos} shows two solutions with four position variables. The residuated triple at the top represents an infixation connective $A\backslash_{3a} C$ and a circumfixion connective $C/_{3a} B$. Note that since this last connective is represented by the pair of white rectangles, it positions itself `around' the $B$ formula. The infixation operation corresponds, at the string level, to the adjoining operation of tree adjoining grammars \parencite{tags} and to the simplest version of the discontinuous connectives of \textcite{mvf11displacement}.

Given the concatenation operation, we can obtain its residuals by plugging them in the Application/Co-Application principles and adding the required quantifiers to make them derivable. However, the general principle is very simple and we can `read off' the definitions directly (although the reader is invited to verify that all the Application/Co-Application principles hold). For the topmost residuated triple this gives the following definition (this connective is labeled $3a$ to indicate it is the first connective with 3 string positions).
\begin{align*}
\| A\bullet_{3a} B \|^{x_0,x_3\phantom{,x_1,x_2}} & = \exists x_1,x_2. [ \| A \|^{x_0,x_1,x_2,x_3} \otimes \| B \|^{x_1,x_2} ] \\	
\| A \backslash_{3a} C \|^{x_1,x_2\phantom{,x_1,x_2}} & = \forall x_0,x_3. [ \| A \|^{x_0,x_1,x_2,x_3} \multimap \| C \|^{x_0,x_3} ] \\
\| C/_{3a} B \|^{x_0,x_1,x_2,x_3} &= \phantom{\forall x_0,x_3.[} \| B \|^{x_1,x_2} \multimap \| C \|^{x_0,x_3} \\
\end{align*}
We can see that the patterns are very similar to the translation of the Lambek calculus connectives: the variables shared between $A$ and $B$ (in the current case $x_1$ and $x_2$) are quantified existentially for the $A\otimes B$ case, the variables shared between $A$ and $C$ are quantified universally  for the $A\multimap C$ case ($x_1$ and $x_2$ here), and the variables shared between $B$ and $C$ (none for this case) are quantified universally for the $B\multimap C$ case. In total each variable is quantified in exactly one of the translation cases.

The residuated triple at the bottom of Figure~\ref{fig:fourpos} assigns positions $x_0,x_1$ to its $A$ formula and positions $x_2,x_3$ to its $B$ formula. In this case, the positions assigned to $A
\otimes  B$ are underdetermined: we can say that nothing is known about the relation between $x_1$ and $x_2$, or between $x_0$ and $x_3$. This case therefore explicitly requires an additional partial order constraint to be a residuated triple. The recursive definitions are as follows.
\begin{align*}
\| A\bullet_{3b} B \|^{x_0,x_1,x_2,x_3} & = \phantom{\exists x_1,x_2. [} \| A \|^{x_0,x_1} \otimes \| B \|^{x_2,x_3} \\	
\| A \backslash_{3b} C \|^{x_2,x_3\phantom{,x_1,x_2}} & = \forall x_0,x_1. [ \| A \|^{x_0,x_1} \multimap \| C \|^{x_0,x_1,x_2,x_3}  ] \\
\| C/_{3b} B \|^{x_0,x_1\phantom{,x_1,x_2}} &= \forall x_2,x_3. [\| B \|^{x_2,x_3} \multimap \| C \|^{x_0,x_1,x_2,x_3} ] \\
\end{align*}
The key case is $A\bullet_{3b} B$, where there would be a loss of information in the information passed to the two subformulas without the additional constraint the $x_1\leq x_2$. 

Now it may seem that this connective is just a formal curiosity. However, it is essentially this pattern, notably the $A \backslash_{3b} C$ connective, which figures in the analysis of the well-known crossed dependencies for Dutch verb clusters of \textcite{mvf11displacement}.

\section{The General Case}

Given linear order of the string position variables, each additional string variable increases the number of possible connectives. We have seen the case for three position variables (the Lambek calculus connectives) and the two residuated triples for four position variables. Are these the only possibilities? And, more generally, how many residuated connectives exist for $k$ position variables.

We want our residuated triples to combine two sequences of components, one containing elementary segments labeled $a$ (corresponding to the left residual) and the other containing elementary segments labeled $b$ (corresponding to the right residual) while allowing an `empty' component between two other components (but not at the beginning or end of a generalised concatenation). Residuated triples can use the `empty' segment $\one$, which corresponds to a sort of placeholder or hole for another segment.
\begin{enumerate}
\item the first segment must be $a$ (concatenations with $b$ as first segment are obtained by left-right symmetry of the residuated triple),
\item there can be no consecutive $a$ segments (that it, if two $a$ segments have already been concatenated, we `lose' the internal structure),
\item for the same reasons, there can be no consecutive $b$ segments,
\item consecutive $\one$ segments do not increase expressivity and are therefore excluded,
\item there must be at least one $b$ segment,
\item the last segment cannot be $\one$ (and, as a consequence of item~1, neither can the first segment). 	
\end{enumerate}

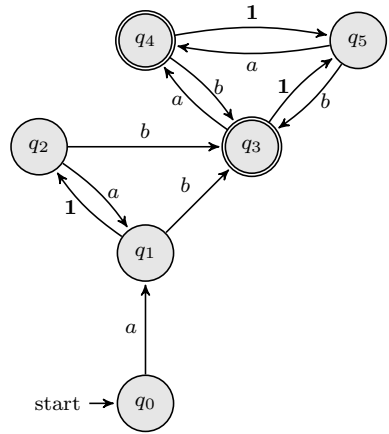
\begin{figure}
\begin{center}
\begin{tikzpicture}[->,>=stealth',shorten >=1pt,auto,node distance=2.0cm,
                    semithick]
 \tikzstyle{every state}=[fill=gray!20]
 \node[state,initial] (zero) {$q_0$};
 \node[state]  (A) [above of=zero] {$q_1$};
 \node[state]  (B) [above left of=A] {$q_2$};
 \node[state,accepting]  (C) [above right of=A] {$q_3$};
 \node[state,accepting]  (D) [above left of=C] {$q_4$};
 \node[state]  (E) [above right of=C] {$q_5$};
 \path (zero) edge node {$a$} (A);
 \path (A) edge [bend left=10, left] node {$\one$} (B); 
 \path (B) edge [bend left=10, right] node {$a$} (A); 
 \path (A) edge node {$b$} (C);
 \path (B) edge node {$b$} (C);
 \path (D) edge [bend left=10, right] node {$b$} (C); 
 \path (C) edge [bend left=10, left] node {$a$} (D); 
 \path (E) edge [bend left=10, right] node {$b$} (C); 
 \path (C) edge [bend left=10, left] node {$\one$} (E); 
 \path (E) edge [bend left=10, below] node {$a$} (D); 
 \path (D) edge [bend left=10, above] node {$\one$} (E); 
\end{tikzpicture}	
\end{center}

\caption{Finite state automaton of concatenation-like operations.}
\label{fig:fsa}	
\end{figure}

The finite state automaton shown in Figure~\ref{fig:fsa} generates all strings which satisfy these requirements. From the start state $q_0$, the only valid symbol is $a$. The condition that we cannot repeat the last symbol then ensures that the states where the last symbol was $a$ (states $q_1$ and $q_4$) can only continue with a $\one$ or a $b$ symbol. Similarly, the states where the last symbol was $\one$ (states $q_2$ and $q_5$) can only continue with an $a$ or a $b$ symbol, and the state where the last symbol was $b$ (state $q_3$) can only continue with $\one$ or $a$. Finally, the states $q_3$, $q_4$ and $q_5$ denote the states where we have seen at least one $b$ symbol. These are accepting states except for $q_5$ (because its last symbol is $\one$).  

We can now show that this machine generates only one two-symbol string $ab$ (corresponding to three string positions and to the simple concatenation of $a$ and $b$) and two three-symbol strings (with four string positions, namely $a\one b$ and $aba$). 

\newcommand{\myskip}{\\[2mm]}
\begin{table}
\begin{center}
\textbf{Two segments, three variables} \myskip
\begin{tikzpicture}
\node at (20em,1.6em) {$x_0$};
\node at (24em,1.6em) {$x_1$};
\node at (28em,1.6em) {$x_2$};
\draw (20em,1em) -- (20em,1.2em);
\draw (24em,1em) -- (24em,1.2em);
\draw (28em,1em) -- (28em,1.2em);
\draw (20em,0em) -- (20em,1em) -- (24em,1em) -- (24em,0em) -- cycle;
\draw [fill=lightgray] (24em,0em) -- (24em,1em) -- (28em,1em) -- (28em,0em) -- cycle; 	
\end{tikzpicture} \myskip
\textbf{Three segments, four variables} \myskip
\begin{tikzpicture}	
\node at (17em,0.65em) {$(3a)$};	
\node at (20em,1.6em) {$x_0$};
\node at (24em,1.6em) {$x_1$};
\node at (28em,1.6em) {$x_2$};
\node at (32em,1.6em) {$x_3$};
\draw (20em,1em) -- (20em,1.2em);
\draw (24em,1em) -- (24em,1.2em);
\draw (28em,1em) -- (28em,1.2em);
\draw (20em,0em) -- (20em,1em) -- (24em,1em) -- (24em,0em) -- cycle;
\draw [fill=lightgray] (24em,0em) -- (24em,1em) -- (28em,1em) -- (28em,0em) -- cycle; 	
\draw (28em,0em) -- (28em,1em) -- (32em,1em) -- (32em,0em) -- cycle;
\end{tikzpicture} \myskip
\begin{tikzpicture}
\node at (17em,0.65em) {$(3b)$};	
\node at (20em,1.6em) {$x_0$};
\node at (24em,1.6em) {$x_1$};
\node at (28em,1.6em) {$x_2$};
\node at (32em,1.6em) {$x_3$};
\draw (20em,1em) -- (20em,1.2em);
\draw (24em,1em) -- (24em,1.2em);
\draw (28em,1em) -- (28em,1.2em);
\draw (20em,0em) -- (20em,1em) -- (24em,1em) -- (24em,0em) -- cycle;
\draw [fill=lightgray] (28em,0em) -- (28em,1em) -- (32em,1em) -- (32em,0em) -- cycle;
\end{tikzpicture} \myskip
\textbf{Four segments, five variables} \myskip
\begin{tikzpicture}	
\node at (17em,0.65em) {$(4a)$};	
\node at (20em,1.6em) {$x_0$};
\node at (24em,1.6em) {$x_1$};
\node at (28em,1.6em) {$x_2$};
\node at (32em,1.6em) {$x_3$};
\node at (36em,1.6em) {$x_4$};
\draw (20em,1em) -- (20em,1.2em);
\draw (24em,1em) -- (24em,1.2em);
\draw (28em,1em) -- (28em,1.2em);
\draw (20em,0em) -- (20em,1em) -- (24em,1em) -- (24em,0em) -- cycle;
\draw [fill=lightgray] (24em,0em) -- (24em,1em) -- (28em,1em) -- (28em,0em) -- cycle; 	
\draw (28em,0em) -- (28em,1em) -- (32em,1em) -- (32em,0em) -- cycle;
\draw [fill=lightgray] (32em,0em) -- (32em,1em) -- (36em,1em) -- (36em,0em) -- cycle; 	
\end{tikzpicture} \myskip
\begin{tikzpicture}	
\node at (17em,0.65em) {$(4b)$};	
\node at (20em,1.6em) {$x_0$};
\node at (24em,1.6em) {$x_1$};
\node at (28em,1.6em) {$x_2$};
\node at (32em,1.6em) {$x_3$};
\node at (36em,1.6em) {$x_4$};
\draw (20em,1em) -- (20em,1.2em);
\draw (24em,1em) -- (24em,1.2em);
\draw (28em,1em) -- (28em,1.2em);
\draw (20em,0em) -- (20em,1em) -- (24em,1em) -- (24em,0em) -- cycle;
\draw [fill=lightgray] (24em,0em) -- (24em,1em) -- (28em,1em) -- (28em,0em) -- cycle; 	
\draw (32em,0em) -- (32em,1em) -- (36em,1em) -- (36em,0em) -- cycle; 	
\end{tikzpicture} \myskip
\begin{tikzpicture}	
\node at (17em,0.65em) {$(4c)$};	
\node at (20em,1.6em) {$x_0$};
\node at (24em,1.6em) {$x_1$};
\node at (28em,1.6em) {$x_2$};
\node at (32em,1.6em) {$x_3$};
\node at (36em,1.6em) {$x_4$};
\draw (20em,1em) -- (20em,1.2em);
\draw (24em,1em) -- (24em,1.2em);
\draw (28em,1em) -- (28em,1.2em);
\draw (20em,0em) -- (20em,1em) -- (24em,1em) -- (24em,0em) -- cycle;
\draw [fill=lightgray] (24em,0em) -- (24em,1em) -- (28em,1em) -- (28em,0em) -- cycle; 	
\draw [fill=lightgray] (32em,0em) -- (32em,1em) -- (36em,1em) -- (36em,0em) -- cycle; 	
\end{tikzpicture} \myskip
\begin{tikzpicture}	
\node at (17em,0.65em) {$(4d)$};	
\node at (20em,1.6em) {$x_0$};
\node at (24em,1.6em) {$x_1$};
\node at (28em,1.6em) {$x_2$};
\node at (32em,1.6em) {$x_3$};
\node at (36em,1.6em) {$x_4$};
\draw (20em,1em) -- (20em,1.2em);
\draw (24em,1em) -- (24em,1.2em);
\draw (28em,1em) -- (28em,1.2em);
\draw (20em,0em) -- (20em,1em) -- (24em,1em) -- (24em,0em) -- cycle;
\draw (28em,0em) -- (28em,1em) -- (32em,1em) -- (32em,0em) -- cycle;
\draw [fill=lightgray] (32em,0em) -- (32em,1em) -- (36em,1em) -- (36em,0em) -- cycle; 	
\end{tikzpicture} \myskip
\begin{tikzpicture}
\node at (17em,0.65em) {$(4e)$};	
\node at (20em,1.6em) {$x_0$};
\node at (24em,1.6em) {$x_1$};
\node at (28em,1.6em) {$x_2$};
\node at (32em,1.6em) {$x_3$};
\node at (36em,1.6em) {$x_4$};
\draw (20em,1em) -- (20em,1.2em);
\draw (24em,1em) -- (24em,1.2em);
\draw (28em,1em) -- (28em,1.2em);
\draw (20em,0em) -- (20em,1em) -- (24em,1em) -- (24em,0em) -- cycle;
\draw [fill=lightgray] (28em,0em) -- (28em,1em) -- (32em,1em) -- (32em,0em) -- cycle;
\draw (32em,0em) -- (32em,1em) -- (36em,1em) -- (36em,0em) -- cycle; 	
\end{tikzpicture} \myskip
\end{center}
	\caption{Concatenation-like operations for two to four string segments}
	\label{tab:conctwofour}
\end{table}

Table~\ref{tab:conctwofour} shows the concatenation-like operations definable with two, three, and four total string segments. The $a$ segments correspond to empty rectangles, the $b$ segments to filled rectangles and the $\one$ segments to empty spaces between the other segments. We can read off the free variables and their linear order for each of the subformulas of a residuated triple. 

For example, the $A$ (and $B\multimap C$) segments of the first item with four segments corresponds to a formula with free variable $x_0,x_1,x_2,x_3$ (in that linear order) whereas the $B$ (and $A\multimap C$) formula corresponds to a formula with free variables $x_1,x_2,x_3,x_4$. Finally, the result of the concatenation formula $C$ (and $A\otimes B$) corresponds to variables $x_0,x_4$, with three separate concatenation operations. We concatenate $a\one a$ to $b\one b$ to produce $abab$. The number of variables shared by the left branch $A$ and the right branch $B$ corresponds to the number of concatenations of elementary segments. If we name this residuated triple $4a$, its recursive definition is as follows.
\begin{align*}
\| A \bullet_{4a} B \|^{x_0,x_4} & = \exists x_1,x_2,x_3. [ \| A \|^{x_0,x_1,x_2,x_3} \otimes \| B \|^{x_1,x_2,x_3,x_4} ] \\
\| A\backslash_{4a} C \|^{x_1,x_2,x_3,x_4} & = \phantom{,x_1,x_2} \forall x_0. [ \| A \|^{x_0,x_1,x_2,x_3} \multimap	 \| C \|^{x_0,x_4} ] \\
\| C/_{4a} B \|^{x_0,x_1,x_2,x_3} &= \phantom{,x_1,x_2} \forall x_4. [ \| B \|^{x_1,x_2,x_3,x_4} \multimap \| C \|^{x_0,x_4}
\end{align*}


As another example, the fourth item with four segments (and five variables) assign the $A$ (and $B\multimap C$) segments the sequence of variables $x_0,x_1,x_2,x_3$, the $B$ and (and $A\multimap C$) formula the variables $x_3,x_4$, and the $C$ (and $A\otimes B$ formula) the variables $x_0,x_1,x_2,x_4$. If we name this residuated triple $4d$, we obtain the following recursive definitions.
\begin{align*}
\| A \bullet_{4d} B \|^{x_0,x_1,x_2,x_4} & = \phantom{,x_1,x_2} \exists x_3. [ \| A \|^{x_0,x_1,x_2,x_3} \otimes \| B \|^{x_3,x_4}] \\	
\| A\backslash_{4d} C \|^{x_3,x_4\phantom{,x_0,x_1}} & = \forall x_0,x_1,x_2. [ \| A \|^{x_0,x_1,x_2,x_3} \multimap \| C \|^{x_0,x_1,x_2,x_4} ] \\
\| C /_{4d} B \|^{x_0,x_1,x_2,x_3} & = \phantom{,x_1,x_2} \forall x_4. [ \| B \|^{x_3,x_4} \multimap   \| C \|^{x_0,x_1,x_2,x_4} ]
 \end{align*}

\begin{table}
\begin{center}
\textbf{Five segments, six variables} \myskip
\begin{tikzpicture}
\node at (17em,0.65em) {$(5a)$};	
\node at (20em,1.6em) {$x_0$};
\node at (24em,1.6em) {$x_1$};
\node at (28em,1.6em) {$x_2$};
\node at (32em,1.6em) {$x_3$};
\node at (36em,1.6em) {$x_4$};
\node at (40em,1.6em) {$x_5$};
\draw (20em,1em) -- (20em,1.2em);
\draw (24em,1em) -- (24em,1.2em);
\draw (28em,1em) -- (28em,1.2em);
\draw (20em,0em) -- (20em,1em) -- (24em,1em) -- (24em,0em) -- cycle;
\draw [fill=lightgray] (24em,0em) -- (24em,1em) -- (28em,1em) -- (28em,0em) -- cycle; 	
\draw (28em,0em) -- (28em,1em) -- (32em,1em) -- (32em,0em) -- cycle;
\draw [fill=lightgray] (32em,0em) -- (32em,1em) -- (36em,1em) -- (36em,0em) -- cycle; 	
\draw (36em,0em) -- (36em,1em) -- (40em,1em) -- (40em,0em) -- cycle;
\end{tikzpicture} \myskip	
\begin{tikzpicture}	
\node at (17em,0.65em) {$(5b)$};	
\node at (20em,1.6em) {$x_0$};
\node at (24em,1.6em) {$x_1$};
\node at (28em,1.6em) {$x_2$};
\node at (32em,1.6em) {$x_3$};
\node at (36em,1.6em) {$x_4$};
\node at (40em,1.6em) {$x_5$};
\draw (20em,1em) -- (20em,1.2em);
\draw (24em,1em) -- (24em,1.2em);
\draw (28em,1em) -- (28em,1.2em);
\draw (20em,0em) -- (20em,1em) -- (24em,1em) -- (24em,0em) -- cycle;
\draw [fill=lightgray] (24em,0em) -- (24em,1em) -- (28em,1em) -- (28em,0em) -- cycle; 	
\draw (28em,0em) -- (28em,1em) -- (32em,1em) -- (32em,0em) -- cycle;
\draw (36em,0em) -- (36em,1em) -- (40em,1em) -- (40em,0em) -- cycle;
\end{tikzpicture} \myskip	
\begin{tikzpicture}	
\node at (17em,0.65em) {$(5c)$};	
\node at (20em,1.6em) {$x_0$};
\node at (24em,1.6em) {$x_1$};
\node at (28em,1.6em) {$x_2$};
\node at (32em,1.6em) {$x_3$};
\node at (36em,1.6em) {$x_4$};
\node at (40em,1.6em) {$x_5$};
\draw (20em,1em) -- (20em,1.2em);
\draw (24em,1em) -- (24em,1.2em);
\draw (28em,1em) -- (28em,1.2em);
\draw (20em,0em) -- (20em,1em) -- (24em,1em) -- (24em,0em) -- cycle;
\draw [fill=lightgray] (24em,0em) -- (24em,1em) -- (28em,1em) -- (28em,0em) -- cycle; 	
\draw (28em,0em) -- (28em,1em) -- (32em,1em) -- (32em,0em) -- cycle;
\draw [fill=lightgray] (36em,0em) -- (36em,1em) -- (40em,1em) -- (40em,0em) -- cycle;
\end{tikzpicture} \myskip	
\begin{tikzpicture}	
\node at (17em,0.65em) {$(5d)$};	
\node at (20em,1.6em) {$x_0$};
\node at (24em,1.6em) {$x_1$};
\node at (28em,1.6em) {$x_2$};
\node at (32em,1.6em) {$x_3$};
\node at (36em,1.6em) {$x_4$};
\node at (40em,1.6em) {$x_5$};
\draw (20em,1em) -- (20em,1.2em);
\draw (24em,1em) -- (24em,1.2em);
\draw (28em,1em) -- (28em,1.2em);
\draw (20em,0em) -- (20em,1em) -- (24em,1em) -- (24em,0em) -- cycle;
\draw [fill=lightgray] (24em,0em) -- (24em,1em) -- (28em,1em) -- (28em,0em) -- cycle; 	
\draw  (32em,0em) -- (32em,1em) -- (36em,1em) -- (36em,0em) -- cycle; 	
\draw [fill=lightgray] (36em,0em) -- (36em,1em) -- (40em,1em) -- (40em,0em) -- cycle;
\end{tikzpicture} \myskip	
\begin{tikzpicture}	
\node at (17em,0.65em) {$(5e)$};	
\node at (20em,1.6em) {$x_0$};
\node at (24em,1.6em) {$x_1$};
\node at (28em,1.6em) {$x_2$};
\node at (32em,1.6em) {$x_3$};
\node at (36em,1.6em) {$x_4$};
\node at (40em,1.6em) {$x_5$};
\draw (20em,1em) -- (20em,1.2em);
\draw (24em,1em) -- (24em,1.2em);
\draw (28em,1em) -- (28em,1.2em);
\draw (20em,0em) -- (20em,1em) -- (24em,1em) -- (24em,0em) -- cycle;
\draw [fill=lightgray] (24em,0em) -- (24em,1em) -- (28em,1em) -- (28em,0em) -- cycle; 	
\draw [fill=lightgray] (32em,0em) -- (32em,1em) -- (36em,1em) -- (36em,0em) -- cycle; 	
\draw (36em,0em) -- (36em,1em) -- (40em,1em) -- (40em,0em) -- cycle;
\end{tikzpicture} \myskip	
\begin{tikzpicture}
\node at (17em,0.65em) {$(5f)$};	
\node at (20em,1.6em) {$x_0$};
\node at (24em,1.6em) {$x_1$};
\node at (28em,1.6em) {$x_2$};
\node at (32em,1.6em) {$x_3$};
\node at (36em,1.6em) {$x_4$};
\node at (40em,1.6em) {$x_5$};
\draw (20em,1em) -- (20em,1.2em);
\draw (24em,1em) -- (24em,1.2em);
\draw (28em,1em) -- (28em,1.2em);
\draw (20em,0em) -- (20em,1em) -- (24em,1em) -- (24em,0em) -- cycle;
\draw (28em,0em) -- (28em,1em) -- (32em,1em) -- (32em,0em) -- cycle;
\draw [fill=lightgray] (32em,0em) -- (32em,1em) -- (36em,1em) -- (36em,0em) -- cycle; 	
\draw (36em,0em) -- (36em,1em) -- (40em,1em) -- (40em,0em) -- cycle;
\end{tikzpicture} \myskip	
\begin{tikzpicture}
\node at (17em,0.65em) {$(5g)$};	
\node at (20em,1.6em) {$x_0$};
\node at (24em,1.6em) {$x_1$};
\node at (28em,1.6em) {$x_2$};
\node at (32em,1.6em) {$x_3$};
\node at (36em,1.6em) {$x_4$};
\node at (40em,1.6em) {$x_5$};
\draw (20em,1em) -- (20em,1.2em);
\draw (24em,1em) -- (24em,1.2em);
\draw (28em,1em) -- (28em,1.2em);
\draw (20em,0em) -- (20em,1em) -- (24em,1em) -- (24em,0em) -- cycle;
\draw (28em,0em) -- (28em,1em) -- (32em,1em) -- (32em,0em) -- cycle;
\draw [fill=lightgray] (36em,0em) -- (36em,1em) -- (40em,1em) -- (40em,0em) -- cycle;
\end{tikzpicture} \myskip	
\begin{tikzpicture}	
\node at (17em,0.65em) {$(5h)$};	
\node at (20em,1.6em) {$x_0$};
\node at (24em,1.6em) {$x_1$};
\node at (28em,1.6em) {$x_2$};
\node at (32em,1.6em) {$x_3$};
\node at (36em,1.6em) {$x_4$};
\node at (40em,1.6em) {$x_5$};
\draw (20em,1em) -- (20em,1.2em);
\draw (24em,1em) -- (24em,1.2em);
\draw (28em,1em) -- (28em,1.2em);
\draw (20em,0em) -- (20em,1em) -- (24em,1em) -- (24em,0em) -- cycle;
\draw [fill=lightgray](28em,0em) -- (28em,1em) -- (32em,1em) -- (32em,0em) -- cycle;
\draw (32em,0em) -- (32em,1em) -- (36em,1em) -- (36em,0em) -- cycle; 	
\draw [fill=lightgray](36em,0em) -- (36em,1em) -- (40em,1em) -- (40em,0em) -- cycle;
\end{tikzpicture} \myskip	
\begin{tikzpicture}	
\node at (17em,0.65em) {$(5i)$};	
\node at (20em,1.6em) {$x_0$};
\node at (24em,1.6em) {$x_1$};
\node at (28em,1.6em) {$x_2$};
\node at (32em,1.6em) {$x_3$};
\node at (36em,1.6em) {$x_4$};
\node at (40em,1.6em) {$x_5$};
\draw (20em,1em) -- (20em,1.2em);
\draw (24em,1em) -- (24em,1.2em);
\draw (28em,1em) -- (28em,1.2em);
\draw (20em,0em) -- (20em,1em) -- (24em,1em) -- (24em,0em) -- cycle;
\draw [fill=lightgray](28em,0em) -- (28em,1em) -- (32em,1em) -- (32em,0em) -- cycle;
\draw (36em,0em) -- (36em,1em) -- (40em,1em) -- (40em,0em) -- cycle;
\end{tikzpicture} \myskip	
\begin{tikzpicture}	
\node at (17em,0.65em) {$(5j)$};	
\node at (20em,1.6em) {$x_0$};
\node at (24em,1.6em) {$x_1$};
\node at (28em,1.6em) {$x_2$};
\node at (32em,1.6em) {$x_3$};
\node at (36em,1.6em) {$x_4$};
\node at (40em,1.6em) {$x_5$};
\draw (20em,1em) -- (20em,1.2em);
\draw (24em,1em) -- (24em,1.2em);
\draw (28em,1em) -- (28em,1.2em);
\draw (20em,0em) -- (20em,1em) -- (24em,1em) -- (24em,0em) -- cycle;
\draw [fill=lightgray](28em,0em) -- (28em,1em) -- (32em,1em) -- (32em,0em) -- cycle;
\draw [fill=lightgray](36em,0em) -- (36em,1em) -- (40em,1em) -- (40em,0em) -- cycle;
\end{tikzpicture} \myskip	
\end{center}
\caption{Concatenation-like operations for five string segments}	
\label{tab:five}
\end{table}

Table~\ref{tab:five} shows the concatenation-like operations definable with five string segments. We give an example of only one of these, because it illustrates a new pattern. As we have seen, some concatenation-like operations require additional order constraints to uniquely define a linear order, for each subformula, on all variables occurring exactly once in this subformula. This was the case for the second possibility with three segments, where we could not infer the order between the $A$ segment $x_0,x_1$ and the $B$ segment $x_2,x_3$ without explicitly requiring $x_1 \leq x_2$.

The second item of Table~\ref{tab:five}, $5b$, shows a different type of underdetermination. When we give the translation of the table entry into a residuated triple $5b$, we obtain the following.
\begin{align*}
\| A\bullet_{5b} B \|^{x_0,x_3,x_4,x_5\phantom{,x_1,x_2}} & = \phantom{,x_0,x_3}\exists x_1,x_2. [ \| A \|^{x_0,x_1,x_2,x_3,x_4,x_5} \otimes \| B \|^{x_1,x_2}	] \\
\| A\backslash_{5b} C \|^{x_1,x_2\phantom{,x_0,x_3,x_4,x_5}} &= \forall x_0,x_3,x_4,x_5. [ \| A \|^{x_0,x_1,x_2,x_3,x_4,x_5} \multimap \| C\|^{x_0,x_3,x_4,x_5}] \\
\| C/_{5b} B \|^{x_0,x_1,x_2,x_3,x_4,x_5} &= \phantom{\forall x_0,x_3,x_4,x_5. [}\| B \|^{x_1,x_2} \multimap \| C\|^{x_0,x_3,x_4,x_5}
\end{align*}

The problematic connective here is $C/_{5b} B$. The order information of its subformulas $B$ and $C$ does not allow us to unambiguously reconstruct the full order: it is compatible with an alternative linear order $x_0,x_3,x_4,x_1,x_2,x_5$, which is the sixth entry $5f$ in Table~\ref{tab:five}. The left residuals of $5b$ and $5f$ cannot be distinguished without an explicit constraint on the linear order for the left residual. In the case above, we need to explicitly state that $x_0 \leq x_1$ and $x_2 \leq x_3$ (technically, since $x_0$ is the leftmost element of the triple, the first constraint is superfluous).

\subsection{How Many Residuated Connectives Are There for Concatenation-Like Operations?}

Since the finite state automaton of Figure~\ref{fig:fsa} is deterministic, each transition produces a symbol and it is therefore easy to use the automaton to enumerate the number of strings\footnote{In the literature on finite state automata it is common to refer to sequences of symbols produced by such an automaton as ``words''. However, we reserve ``words'' to refer to elements in the lexicon of a type-logical grammar and exclusively use ``string'' for a sequence of symbols produced by a finite state automaton.} of a certain length $k$.
 


We can also use the machine to directly compute the number of words, either by using a standard dynamic programming approach or by solving the linear recurrence specified by the automaton to produce a closed form.  For example, there is a single length 1 path to $q_1$ (the path from the start state $q_0$). For paths of length greater than 1, the number of paths to $q_1$ of length $k$ is equal to the number of paths of length $k-1$ to $q_2$. In general, the number of paths of length $k$ to a state is the sum of the paths of length $k-1$ which can reach this state in one step. Writing out the full definition then gives the following set of linear recurrences, where $p[Q][K]$ denotes the number of paths of length $K$ which reach state $Q$. In addition, $p[k]$ denotes the number of accepting paths of length $k$ and it is the sum of the number of paths to the two accepting states $q_3$ and $q_4$. 
\begin{align*}
p[q_1][1] & = 1 \\
p[q_1][k] & = p[q_2][k-1] \\
p[q_2][k] & = p[q_1][k-1] \\
p[q_3][k] &= p[q_4][k-1] + p[q_5][k-1] + p[q_1][k-1] + p[q_2][k-1] \\
p[q_4][k] &= p[q_3][k-1] + p[q_5][k-1] \\
p[q_5][k] &= p[q_3][k-1] + p[q_4][k-1] \\
p[k] &= p[q_3][k] + p[q_4][k] 
\end{align*}

We can simplify these equations by observing that for each $k$ there is exactly one path arriving at $q$ in $k$ steps from either $q_2$ (for $k-1$ even) or $q_1$ (for $k-1$ odd). So we can simplify $p[q_1][k-1] + p[q_2][k-1]$ to $1$. In addition, because of the symmetries in the automaton, there are exactly as many paths reaching $q_4$ as there are reaching $q_5$ for any $k$, so we can replace $p[q_5][k]$ by $p[q_4][k]$ without changing the results. This simplifies the equations as follows.
\begin{align*}
p[q_3][0] = p[q_3][1] &= 0 \\
p[q_4][0] = p[q_4][1] = p[q_4][2] &= 0 \\
p[q_3][k] &= 2*p[q_4][k-1] + 1  & (k > 1)\\
p[q_4][k] &= p[q_3][k-1] + p[q_4][k-1] &(k> 2)\\
p[k] &= p[q_3][k] + p[q_4][k] 
\end{align*}
We can now show the following.
\begin{align}
\label{eq:odd}	p[q_3][k] &= p[q_4][k] & (\textit{for}\  k\  \textit{odd}) \\
\label{eq:even}	p[q_3][k] &= p[q_4][k]+1 & (\textit{for}\  k \ \textit{even and} \, \geq 2)
\end{align}

This is an easy induction: it is trivially true for $k=1$. Now assume Equations~\ref{eq:odd} and~\ref{eq:even} hold for all $k' < k$. 

If $k$ is even, $k-1$ is odd, and induction hypothesis gives us $p[q_3][k-1] = p[q_4][k-1]$ and we need to show that $p[q_3][k]= p[q_4][k] + 1$, given $k\geq 2$. Using $p[q_3][k-1] = p[q_4][k-1]$, we can simplify $p[q_4][k]= p[k_3][k-1] + p[p4][k-1]$ to $p[q_4][k]= 2*p[q_4][k]$. But since $p[q_3][k]= 
2*p[q_4][k]+1$ we have therefore shown that $p[q_3][k]= p[q_4][k] + 1$.  

If $k$ is odd, $k-1$ is even, and induction hypothesis gives us $p[q_3][k-1] = p[q_4][k-1]+1$. We have already verified $k=1$, so we only need to verify $k\geq 3$. Again, using $p[q_3][k-1] = p[q_4][k-1]+1$ to substitute $p[q_4][k-1]+1$ for $p[q_3][k-1]$ in the equation for $p[q_4][k]$ produces $p[q_4][k] = 2*p[q_4][k-1] + 1$ and we have therefore shown that $p[q_3][k]=p[q_4][k]$ as required. 

We can use Equations~\ref{eq:odd} and~\ref{eq:even} to further simplify the machine equations and end up with the following.

For $k$ odd, we have
\begin{align*}
p[q_4][k] = 2*p[q_4][k-1]+1 \\
p[q_3][k] = 2*p[q_3][k-1]-1
\end{align*}
and therefore
\begin{align*}
p[k] &= 2*p[q_4][k-1]+1 + 2*p[q_3][k-1]-1\\
&       = 2*p[k-1]
\end{align*}

For $k$ even and $\geq 2$, we have
\begin{align*}
p[q_4][k] = 2*p[q_4][k-1] \\
p[q_3][k] = 2*p[q_3][k-1]+1
\end{align*}
and therefore
\begin{align*}
p[k] &= 2*p[q_4][k-1] + 2*p[q_3][k-1]+1\\
&       = 2*p[k-1] + 1
\end{align*}

The number of residuated connectives definable in first-order linear logic with partial order constraints therefore corresponds to sequence A000975 of the Online Encyclopedia of Integer Sequences \parencite{oeis}. Giving us the sequence the following sequence of the number of residuated triples
\[
0,1,2,5,10,21,42,85,170,341,682,\ldots
\]
for $1, 2, 3,\ldots$ total string components and for $2,3,4,\ldots$ total string positions\footnote{A closed form solution for this recurrence is the following \parencite{oeis}.\[
p[n] = \left\lceil \frac{2(2^n -1)}{3} \right\rceil
\]}.

\subsection{Well-Nestedness}

One important property often imposed on linguistic formalisms is the property of well-nestedness \parencite{kallmeyer}. In the current context, this means that with respect to the finite state automaton of Figure~\ref{fig:fsa}, we restrict ourselves to those paths where, whenever we encounter an $a$ symbol after a $b$, there can be no further $b$ symbols. In other words, the $b$s are sandwiched between the $a$, but not inversely. The simplest non-wellnested combination is $abab$. 

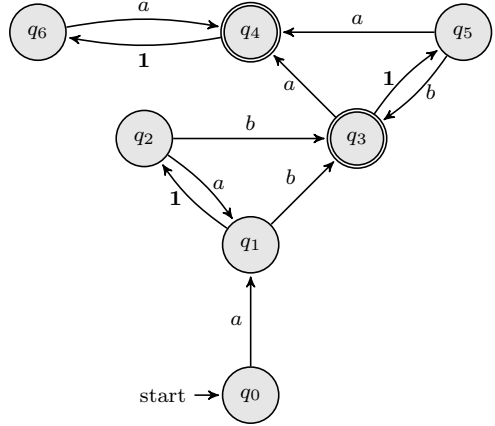
\begin{figure}
\begin{center}
\begin{tikzpicture}[->,>=stealth',shorten >=1pt,auto,node distance=2.0cm,
                    semithick]
 \tikzstyle{every state}=[fill=gray!20]
 \node[state,initial] (zero) {$q_0$};
 \node[state]  (A) [above of=zero] {$q_1$};
 \node[state]  (B) [above left of=A] {$q_2$};
 \node[state,accepting]  (C) [above right of=A] {$q_3$};
 \node[state,accepting]  (D) [above left of=C] {$q_4$};
 \node[state]  (E) [above right of=C] {$q_5$};
 \node[state]  (F) [above left of=B] {$q_6$};
 \path (zero) edge node {$a$} (A);
 \path (A) edge [bend left=10, left] node {$\one$} (B); 
 \path (B) edge [bend left=10, right] node {$a$} (A); 
 \path (A) edge node {$b$} (C);
 \path (B) edge node {$b$} (C);
 %
 \path (C) edge [left] node {$a$} (D); 
\path (E) edge [above] node {$a$} (D); 
 \path (E) edge [bend left=10, right] node {$b$} (C); 
 \path (C) edge [bend left=10, left] node {$\one$} (E); 
 \path (F) edge [bend left=10, above] node {$a$} (D); 
 \path (D) edge [bend left=10, below] node {$\one$} (F); 
\end{tikzpicture}	
\end{center}

\caption{Variant of the finite state automatic of Figure~\ref{fig:fsa} for well-nested operations}
\label{fig:fsawn}	
\end{figure}

We can write out the linear recurrences as before. The number of paths to $q_3$ and $q_5$ are easily established to be the following.
\begin{align*}
p[q_3][2k]\phantom{\rule{0pt}{1ex}+1} &= k  \\
p[q_3][2k+1] &= k \\
p[q_5][2k]\phantom{\rule{0pt}{1ex}+1} &= k - 1  & k > 0\\
p[q_5][2k+1] &= k \\
\end{align*}
Then given that $p[q_6][n] = p[q_4][n-1]$, we can establish the number of paths to $q_4$ as follows.
\begin{align*}
p[q_4][2k]\phantom{\rule{0pt}{1ex}+1} &= q[3][2k-1] + q[5][2k-1] + q[4][2(k-1)] \\
p[q_4][2k+1] &= q[3][2k]\phantom{\rule{0pt}{1ex}-1} + q[5][2k]\phantom{\rule{0pt}{1ex}-1} + q[4][2(k-1)+1] \\
\end{align*}
Simplifying the above recurrence with the calculated values for $q_3$ and $q_5$ produces the following.
\begin{align*}
p[q_4][2k]\phantom{\rule{0pt}{1ex}+1} &= q[4][2(k-1)]\phantom{\rule{0pt}{1ex}+1} + 2(k-1) &= k(k-1)\phantom{\rule{0pt}{1ex}^2} \\
p[q_4][2k+1] &= q[4][2(k-1)+1] + 2k &= k^2\phantom{(k+1)} \\
\end{align*}
The number of paths to a final state of the automaton is then obtain by simply adding the number of paths to $q_3$ to those to $q_4$, which gives us the following solutions after some elementary arithmetic.
\begin{align*}
p[2k]\phantom{\rule{0pt}{1ex}+1} &= p[q_3][2k] + p[q_4][2k] \\
&= k + k(k-1) &= k^2\phantom{(k+1)} \\
 p[2k+1] &= p[q_3][2k+1] + p[q_4][2k+1] \\
 & = k + k^2 &= k(k+1)\phantom{\rule{0pt}{1ex}^2} \\
 \end{align*}

\editout{
\begin{align*}
p[q_1][1] & = 1 \\
p[q_2][k] & = p[q_1][k-1] \\
p[q_3][k] &= p[q_5][k-1] + p[q_1][k-1] + p[q_2][k-1] \\
p[q_4][k] &= p[q_3][k-1] + p[q_6][k-1] \\
p[q_5][k] &= p[q_3][k-1] \\
p[q_6][k] &= p[q_4][k-1] \\
p[k] &= p[q_3][k] + p[q_4][k] 
\end{align*}

\begin{align*}
p[q_3][k] &= p[q_5][k-1] + 1 \\
p[q_4][k] &= p[q_3][k-1] + p[q_5][k-1] + p[q_6][k-1] \\
p[q_5][k] &= p[q_3][k-1] \\
p[q_6][k] &= p[q_4][k-1] \\
p[k] &= p[q_3][k] + p[q_4][k] 
\end{align*}

\begin{align*}
p[q_3][k] &= p[q_3][k-2] + 1 \\
p[q_4][k] &= p[q_3][k-1] + p[q_3][k-2] + p[q_4][k-2] \\
p[k] &= p[q_3][k] + p[q_4][k] 
\end{align*}

\begin{align*}
p[q_3][k] &= \lfloor k / 2 \rfloor \\
p[q_4][k] &= p[q_4][k-1] + \lfloor ((k-1)/2) \rfloor \\
p[q_4][k] &= \lfloor (k-1)/2 \rfloor + \lfloor (k-2)/2 \rfloor + p[q_4][k-2]
\end{align*}

}

An alternative way to state this same solution is the following.
\begin{align*}
p[n] &= \lfloor (n/2) \rfloor * \lceil n/2 \rceil \\
\end{align*}

Accordingly, the number of well-nested residuated connectives is the following
\[
0, 1, 2, 4, 6, 9, 12, 16, 20, 25, 30, 36, \ldots
\]
\noindent for $1, 2, 3, 4, 5, \ldots$ segments and $2, 3, 4, 5, 6, \ldots$ string position variables. This corresponds to sequence A002620 of the Online Encyclopedia  of Integer Sequences \parencite{oeis}.

As a sanity check, we can verify that 4 out of 5 of the four segment possibilities of Table~\ref{tab:conctwofour} are well-nested (only $4a$ is not) whereas 6 out of 10 of the five segment possibilities of Table~\ref{tab:five} are well-nested (the exceptions being $5a$, $5c$, $5d$, and $5h$).

\subsection{Partial Order Constraints in Practice}
\label{sec:poexe}

As an example, we will give an analysis of the sentence `John left before Mary did' based on the analysis of \textcite{mvf11displacement}. We assign `John' and `Mary' the formulas $np(0,1)$ and $np(3,4)$ respectively (based on their positions in the string). We assign `left' the formula $np\backslash s$, which at positions $1,2$ translates to $\forall A. [np(A,1)\multimap s(A,2)]$. We assign the `before' the formula $((np\backslash s)\backslash (np\backslash s))/s$ (that is, it selects a sentence to its right and a $vp = np\backslash s$ to its left to return a $vp$). This translates to the following formula.
\[
\forall B. [ s(3,B) \multimap  \forall D . [ \forall x. [np(x_0,D) \multimap s(x_0,2) ] \multimap \forall C. [np(C,D) \multimap s(C,B)]]]
\]
Finally, the complicated formula is assigned to `did'. In terms of the residuated connectives it is assigned to formula $((vp/_{3a}vp)/vp)\backslash_{4d}(vp/_{3a}vp)$. As a reminder, we restate the relevant translations of the connectives occurring in this formula.
\begin{align*}
\| C/B \|^{x_0,x_1\phantom{,x_2,x_3}} &= \phantom{,x_1,x_2}\forall x_2. [ \|	 B \|^{x_1,x_2} \multimap \| C\|^{x0,x_2} ] \\
\| C/_{3a}B \|^{x_0,x_1,x_2,x_3} &=  \phantom{\forall x_0,x_1,x_2.[} \|	 B \|^{x_1,x_2} \multimap \| C\|^{x0,x_3}  \\
\| A\backslash_{4d}C \|^{x_3,x_4\phantom{,x_2,x_3}} &= \forall x_0,x_1,x_2. [  \| A \|^{x_0,x_1,x_2,x_3} \multimap \| C \|^{x_0,x_1,x_2,x_4} ]
\end{align*}
Given these translations, we can translate this formula into first-order linear logic as follows.
\begin{align*}
	\| ((vp/_{3a}vp)/vp)\backslash_{4d}(vp/_{3a}vp) \|^{4,5} \\
\forall F,I,J.	\| (vp/_{3a}vp)/vp \|^{F,I,J,4} \multimap \| vp/_{3a}vp \|^{F,I,J,5} \\
\forall F,I,J.  [ \forall x_1,  \| vp \|^{4,x_1} \multimap \| vp/_{3a}vp \|^{F,I,J,x_1} ] \multimap \| vp \|^{I,J} \multimap \| vp \|^{F,5} \\
\forall F,I,J.  [ \forall x_1,  \| vp \|^{4,x_1} \multimap \| vp \|^{I,J} \multimap \| vp \|^{F,x_1} ] \multimap \| vp \|^{I,J} \multimap \| vp \|^{F,5} \\
\end{align*}
We have left the final $vp = np\backslash s$ subformulas untranslated. We can see that aside for some fairly complicate manipulation with string positions, to which we will return shortly, the formula simply indicates it select a function of two $vp$'s into a single $vp$ to become a $vp$ modifier. 

%


Given these translations, Figure~\ref{fig:jlbmd} shows the formula unfolding for the sentence `John left before Mary did'. Each node indicates the corresponding linear order on the variables occurring once in this subformula. The complex formula `did' has many branchings but referring back to the position variables allows to to identify which node corresponds to which subformula in the translation. For example, the node labeled $F,I,J,x_1$ corresponds to (the leftmost occurrence of) the formula $vp/_{3a}vp$.

\begin{figure}
\begin{center}
\begin{tikzpicture}
\node (john) at (14em,60em) {$np(0,1)$};
\node (johnl) [above=0.3em of john] {\textit{John}};
\node (mary) at (44em,60em) {$np(3,4)$};
\node (maryl) [above=0.3em of mary] {\textit{Mary}};
\node (left) at (20em,60em) {$1,2$};
\node (ll) [above=0.3em of left] {\textit{left}};
\node (eq) [below=2em of left] {$1,2$};
\draw (eq) -- (left) node [midway] {$A \
  \ \ \ $};
\node (mid) [left=0.66em of eq] {};
\node (ef) [left=0.66em of mid] {$np(A,1)$};
\node (sbot) [below=2em of mid] {$s(A,2)$};
\draw (sbot) -- (eq);
\draw (ef) -- (sbot);
\node (left) at (40em,60em) {$2,3$};
\node (ll) [above=0.3em of left] {\textit{before}};
\node (eq) [below=2em of left] {$2,3$};
\draw (eq) -- (left) node [midway] {$B \
  \ \ \ $};
\node (mid) [left=0.66em of eq] {};
\node (ef) [left=0.66em of mid] {$s(3,B)$};
\node (sbot) [below=2em of mid] {$2,B$};
\draw (sbot) -- (eq);
\draw (ef) -- (sbot);
\node (dq) [below=2em of sbot] {$2,B$};
\draw (sbot) -- (dq) node [midway] {$D \
  \ \ \ $};
\node (ctmp) [left=1em of dq] {};
\node (d2) [left=1em of ctmp] {$D,2$};
\node (vp) [below=2em of ctmp] {$D,B$};
\draw (d2) -- (vp);
\draw (vp) -- (dq);
\node (x3q) [above=2em of d2] {$D,2$};
\draw[<-,semithick,dotted] (d2) -- (x3q)  node [midway] {$x_0\
  \ \ \ $};
\node (ctmp) [left=0.66em of x3q] {};
\node (parc) [right=1.0em of ctmp] {};
\node (nx3i) [left=0.66em of ctmp] {$np(x_0,D)$};
\node (sx3j) [above=2em of ctmp] {$s(x_0,2)$};
\begin{scope}
\path [clip] (nx3i) -- (parc.center) -- (sx3j);
\draw (parc) circle (2.5ex);
\end{scope}
\begin{scope}
\begin{pgfinterruptboundingbox}
\path [clip] (parc) circle (2.5ex) [reverseclip];
\end{pgfinterruptboundingbox}
\draw [semithick,dotted] (parc) -- (nx3i);
\draw [semithick,dotted] (parc) -- (sx3j);
\end{scope}
\node (c) [below=2em of vp] {$D,B$};
\draw (c) -- (vp) node [midway] {$C\
  \ \ \ $};
\node (mid) [left=0.66em of c] {};
\node (np) [left=0.66em of mid] {$np(C,D)$};
\node (s) [below=2em of mid] {$s(C,B)$};
\draw (s) -- (c);
\draw (s) -- (np);
\node (cpos) at (30em,30em) {$\quad$};
\node (lollipc) [right=0.66em of cpos] {$\phantom{(}$};
\node (aib) [right=0.5em of cpos] {$F,I,J,4$};
\node (alollip) [left=0.66em of cpos] {$4,x_1$};
\node (blollip) [above=2.5em of cpos] {$F,I,J,x_1$};
\begin{scope}
\begin{pgfinterruptboundingbox}
\path [clip] (lollipc) circle (2.5ex) [reverseclip];
\end{pgfinterruptboundingbox}
\draw [semithick,dotted] (lollipc) -- (blollip);
\draw [semithick,dotted] (lollipc) -- (alollip);
\end{scope}
\begin{scope}
 \path [clip] (alollip) -- (lollipc.center) -- (blollip);
\draw (lollipc.center) circle (2.5ex);
\end{scope}
\node (x1q) [below=2em of aib] {$F,I,J,4$};
\draw[<-,semithick,dotted] (x1q) -- (aib) node [midway] {$x_1\
  \ \ \ $};
\node (gq) [below=2em of alollip] {$4,x_1$};
\draw (gq) -- (alollip) node [midway] {$G\ \ \ \ \ $};
\node (ctmp) [left=0.66em of gq] {};
\node (gl) [left=0.66em of ctmp] {$np(G,4)$};
\node (gb) [below=2.5em of ctmp] {$s(G,x_1)$};
\draw (gq) -- (gb);
\draw (gl) -- (gb);
\node (fijx1l) [left=-1em of blollip] {};
\node (cpos) [left=1.66em of fijx1l] {};
\node (vpij) [left=5.66em of cpos] {$I,J$};
\node (vpfx1) [above=2em of cpos] {$F,x_1$};
\begin{scope}
\path [clip] (vpij) -- (fijx1l.center) -- (vpfx1);
\draw (fijx1l) circle (2.5ex);
\end{scope}
\begin{scope}
\begin{pgfinterruptboundingbox}
\path [clip] (fijx1l) circle (2.5ex) [reverseclip];
\end{pgfinterruptboundingbox}
\draw [semithick,dotted] (fijx1l) -- (vpij);
\draw [semithick,dotted] (fijx1l) -- (vpfx1);
\end{scope}
\node (gq) [below=2em of vpij] {$4,x_1$};
\draw (gq) -- (vpij) node [midway] {$H\ \ \ \ \ $};
\node (ctmp) [left=0.66em of gq] {};
\node (gl) [left=0.66em of ctmp] {$np(H,I)$};
\node (gb) [below=2.5em of ctmp] {$s(H,J)$};
\draw (gq) -- (gb);
\draw (gl) -- (gb);
\node (x2q) [above=2em of vpfx1] {$F,x_1$};
\draw[<-,semithick,dotted] (vpfx1) -- (x2q)  node [midway] {$x_2\
  \ \ \ $};
\node (ctmp) [left=0.66em of x2q] {};
\node (parc) [right=1.0em of ctmp] {};
\node (nx2f) [left=0.66em of ctmp] {$np(x_2,F)$};
\node (sx2x1) [above=2em of ctmp] {$s(x_2,x_1)$};
\begin{scope}
\path [clip] (nx2f) -- (parc.center) -- (sx2x1);
\draw (parc) circle (2.5ex);
\end{scope}
\begin{scope}
\begin{pgfinterruptboundingbox}
\path [clip] (parc) circle (2.5ex) [reverseclip];
\end{pgfinterruptboundingbox}
\draw [semithick,dotted] (parc) -- (nx2f);
\draw [semithick,dotted] (parc) -- (sx2x1);
\end{scope}
\node (ctmp) [right=0.66em of x1q] {};
\node (45) [right=0.66em of ctmp] {$4,5$};
\node (fij5) [below=2em of ctmp] {$F,I,J,5$};
\draw (x1q) -- (fij5);
\draw (45) -- (fij5);
\node (did45) [above=2em of 45] {$4,5$};
\node (did) [above=0.3em of did45] {\textit{did}};
\draw (did45) -- (45) node [midway] {$\qquad\quad\ F,I,J$};
\node (tmppos) [left=5em of fij5] {};
\node (tmpvp) [below=9em of tmppos] {$I,J$};
\node (x3q) [above=2em of tmpvp] {$I,J$};
\draw[<-,semithick,dotted] (tmpvp) -- (x3q)  node [midway] {$x_3\
  \ \ \ $};
\node (ctmp) [left=0.66em of x3q] {};
\node (parc) [right=1.0em of ctmp] {};
\node (nx3i) [left=0.66em of ctmp] {$np(x_3,I)$};
\node (sx3j) [above=2em of ctmp] {$s(x_3,J)$};
\begin{scope}
\path [clip] (nx3i) -- (parc.center) -- (sx3j);
\draw (parc) circle (2.5ex);
\end{scope}
\begin{scope}
\begin{pgfinterruptboundingbox}
\path [clip] (parc) circle (2.5ex) [reverseclip];
\end{pgfinterruptboundingbox}
\draw [semithick,dotted] (parc) -- (nx3i);
\draw [semithick,dotted] (parc) -- (sx3j);
\node (tmpbot) [right=2em of tmpvp] {};
\node (f5) [below=2em of tmpbot] {$F,5$};
\draw (f5) -- (tmpvp);
\draw (f5) -- (fij5);
\node (eq) [below=2em of f5] {$F,5$};
\draw (eq) -- (f5) node [midway] {$E \
  \ \ \ $};
\node (mid) [left=0.66em of eq] {};
\node (ef) [left=0.66em of mid] {$np(E,F)$};
\node (sbot) [below=2em of mid] {$s(E,5)$};
\draw (sbot) -- (eq);
\draw (ef) -- (sbot);
\end{scope}
\end{tikzpicture}
\end{center}
\caption{Proof structure formed from the formula unfolding for `John left before Mary did'}
\label{fig:jlbmd}
\end{figure}

Table~\ref{tab:matching} shows the possible matchings between positive and negative atomic formulas. The rows of the table represent the choices for the positive formulas, whereas the columns represent the choices for the negative formulas. The positive $s(0,5)$ formula represents the conclusion, the other positive formulas are those which are premisses of their link.  
\begin{table}
\begin{center}
\begin{tabular}{|l|c|c|c|c|c|} \hline 
\cellcolor{gray!20} &  \cellcolor{gray!20} $np(x_0,D)$ & \cellcolor{gray!20} $np(0,1)$ & \cellcolor{gray!20}$np(3,4)$ & \cellcolor{gray!20}$np(x_2,F)$ & \cellcolor{gray!20}$np(x_3,I)$ \\ \hline	
\cellcolor{gray!20}$np(C,D)$ & & & &\cellcolor{gray!80}2 & \\ \hline
\cellcolor{gray!20}$np(A,1)$ & & & & & \cellcolor{gray!80}9 \\ \hline
\cellcolor{gray!20}$np(E,F)$ & &\cellcolor{gray!80}4 & & & \\ \hline
\cellcolor{gray!20}$np(G,4)$ & & &\cellcolor{gray!80}8 & & \\ \hline
\cellcolor{gray!20}$np(H,I)$ &\cellcolor{gray!80}10 & & & & \\ \hline
\end{tabular}
\\

\bigskip

\begin{tabular}{|l|c|c|c|c|c|} \hline 
\cellcolor{gray!20}& \cellcolor{gray!20}$s(A,2)$ & \cellcolor{gray!20}$s(E,5)$ & \cellcolor{gray!20}$s(H,J)$ & \cellcolor{gray!20}$s(G,x_1)$ & \cellcolor{gray!20}$s(C,B)$ \\ \hline	
\cellcolor{gray!20}$s(x_0,2)$ & & & \cellcolor{gray!80}6 & & \\ \hline
\cellcolor{gray!20}$s(3,B)$ & &  & &\cellcolor{gray!80}7 & \\ \hline
\cellcolor{gray!20}$s(x_3,J)$ &\cellcolor{gray!80}5 & & & & \\ \hline
\cellcolor{gray!20}$s(x_2,x_1)$ & & & & & \cellcolor{gray!80}1 \\ \hline
\cellcolor{gray!20}$s(0,5)$ & & \cellcolor{gray!80}3 & & & \\ \hline
\end{tabular}
\end{center}
\caption{Possible axiom connectives for the proof structure in Figure~\ref{fig:jlbmd}, with the columns representing the negative occurrences and the rows the positive ones.}
\label{tab:matching}
\end{table}
Each of the candidate proof structures for the goal sequent is one of the perfect matchings of the positive with the negative formulas. However, since there are $n!$ matchings, brute force search is to be avoided as much as possible. Just for the current example, there are $5! = 120$ choices for the $np$ formulas and the same number of choices for the $s$ formulas. Given that these choices are independent, this amounts to a total of $14.400$ different possible proof structures.

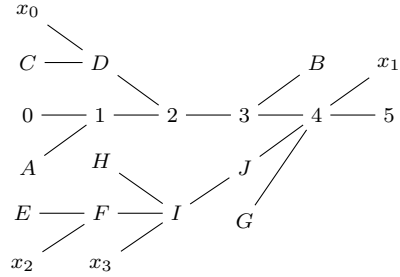
\begin{figure}
\begin{center}
\begin{tikzpicture}
\node (0) at (10em,10em) {$0$};
\node (1) [right=2em of 0] {$1$};
\node (2) [right=2em of 1] {$2$};
\node (3) [right=2em of 2] {$3$};
\node (4) [right=2em of 3] {$4$};
\node (5) [right=2em of 4] {$5$};
\draw (0) -- (1); 	
\draw (1) -- (2); 	
\draw (2) -- (3); 	
\draw (3) -- (4); 	
\draw (4) -- (5); 	
\node (a) [below=1em of 0] {$A$};
\draw (a) -- (1);
\node (d) [above=1em of 1] {$D$};
\draw (d) -- (2);
\node (c) [above=1em of 0] {$C$};
\node (x0) [above=1em of c] {$x_0$};
\draw (c) -- (d);
\draw (x0) -- (d);
\node (j) [below=1em of 3] {$J$};
\draw (j) -- (4);
\node (tmp) [left=2em of j] {};
\node (i) [below=1em of tmp] {$I$};
\draw (i) -- (j);
\node (f) [left=2em of i] {$F$};
\draw (f) -- (i);
\node (e) [left=2em of f] {$E$};
\draw (e) -- (f);
\node (x2) [below=1em of e] {$x_2$};
\draw (x2) -- (f);
\node (x3) [below=1em of f] {$x_3$};
\node (h) [above=1em of f] {$H$};
\draw (x3) -- (i);
\draw (h) -- (i);
\node (g) [below=1em of j] {$G$};
\draw (g) -- (4);
\node (x1) [above=1em of 5] {$x_1$};
\draw (4) -- (x1);
\node (b) [above=1em of 4] {$B$};
\draw (3) -- (b);
\end{tikzpicture}
\end{center}
\caption{The partial order constraints corresponding to the proof structure of Figure~\ref{fig:jlbmd}.}
\label{fig:po}
\end{figure}

Fortunately, there are quite a number of constraints on the possible connections in the proof structure. The partial order constraints are one of those. Figure~\ref{fig:po} summarises the partial order constraints for the structure of Figure~\ref{fig:jlbmd}. 
The partial order constraints allow us to avoid connecting $s(3,B)$ to $s(A,2)$ since it fails both the $3\leq B$ constraint (when unifying $B$ to 2) and the $A\leq 1$ constraint (when unifying $A$ to 3). A slightly less obvious connection which fails the constraint is the connection between $s(x_2,x_1)$ and $s(H,J)$. Here we have  $J \leq 4$, but also $4 < x_1$. Unifying $J$ to $x_1$ would therefore produce the contradicting $x_1 \leq 4$ and $4< x_1$.  

Many potential axioms connections are excluded by a simply failure of unification between the two atoms: the positive atom $s(x_2,x_1)$ cannot connect either to $s(A,2)$ or to $s(E,5)$ (since $x_1$ does not unify with either 2 or 5).

Finally, the contractability condition excludes many other connections. The metavariables $F$, $I$, and $J$ have free occurrences at many nodes. This notably means none of them can unify with $x_1$, $x_2$ or $x_3$ without violating the contraction condition. Similarly, $x_0$ cannot unify with $B$, $C$, or $D$. In general, the eigenvariable of a universal link can never appear on the `wrong' side of its link (the part to which the arrow points), since this would correspond to a violation of the eigenvariable condition in the sequent calculus.

Now, returning to our proof structure, we can see there is only a single possibility for the positive atomic formula $s(x_2,x_1)$. We have already seen that $s(A,2)$ and $s(E,5)$ do no unify and that $s(H,J)$ fails on the partial order constraint. This leaves only $s(C,B)$ and $s(G,x_1)$. However, $s(G,x_1)$ fails on the proof net condition: unifying $G$ to $x_2$ produces an occurrence of $x_2$ on the $4,x_1$ node of the proof structure above the $G$ link (since it is on the existential frontier of $G$). And a reduction of the par link requires an identification of this node with the $F,I,J,x_1$ node, thereby producing an occurrence of $x_2$ on the wrong side of its universal link. Therefore, the only possible connection for $s(x_2,x_1)$ is to $s(C,B)$, unifying $C=x_2$ and $B=x_1$. This fills in the first cell labeled 1 of Table~\ref{tab:matching}. This unification then turns the positive $np(C,D)$ formula into $np(x_2,D)$ which can only unify with $np(x_2,F)$, filling cell 2 of the table.
 
We can now turn to the goal formula $s(0,5)$. Since we have already connected the $s(C,B)$ formula to $s(x_2,x_1)$ this option is no longer available, and the $s(A,2)$ and $s(G,x_1)$ options are excluded by failure of unification. Finally, $s(H,J)$ is excluded because the $J\leq 4$ partial order constraint would contradict unifying $J$ to 5.  This leaves only the $s(E,5)$ possibility, unifying $E$ to 0, as indicated by cell 3 of the table.

After these unifications the negative $np(E,F)$ has become $np(0,D)$ which only unifies with $np(0,1)$, instantiating $D$ to 1, and filling cell 4 of the table. We have now essentially solved the linking problem and the remaining $s$ connections can only be made in a single way, filling cells 5 to 7 in the table. Following that, we can apply similar reasoning to the $np$ connections and fill the remaining cells (cells 8 to 10).

What we have shown is that even a for a quite complex proof structure such as the one in Figure~\ref{fig:jlbmd}, the partial order constraints combined with the proof net conditions can allow us to produce the unique solution while avoiding all backtracking. Given the essentially non-deterministic natural of natural language parsing (sentences can have multiple readings and our parser should therefore produce as many proofs), we will in many cases be required to use some form of backtracking. But this examples gives an illustration of how powerful the combined constraints are.
 
\section{The Empty String}

Up until now, we have not explicitly allowed string segments to be empty. 
However, there are some well-know applications of empty string, notably the treatment of extraction in variants of the Lambek calculus.
We can add a variant of extraction as a residuated pair as follows.
\begin{align*}
\| A \multimap C \|^{y,z} & = \forall x. [\| A \|^{x,x}] \multimap \| C
\|^{y,z} \\
\| A \otimes B \|^{y,z} & = \forall x. [\| A \|^{x,x}] \otimes \|B \|
\end{align*}

Even though this works in many cases, there is a potential problem here: suppose the extracted element is a $vp$, that is the Lambek calculus formula $np\backslash s$, with the standard translation into first-order linear logic of 
\[
\forall x_0. [ np(x_0,x_1) \multimap s(x_0,x_2)]
\] 
\noindent corresponding to a $vp$ at positions $x_1,x_2$. When we plug this formula into the $A$ argument of the implication selecting an empty argument, the result is the identification of $x_1$ and $x_2$, producing the formula
 \[
 \forall x_1 \forall x_0. [ np(x_0,x_1) \multimap s(x_0,x_1)]\]
 \noindent for this extracted $vp$.

Compare this to an extracted formula corresponding to $s/np$. It would be translated into 
\[
\forall x_2. [np(x_1,x_2) \multimap s(x_0,x_2) ]\] \noindent at positions $x_0,x_1$. Turning this into the empty string identifies $x_0$ with $x_1$, producing the following \[\forall x_1\forall x_2. [np(x_1,x_2)\multimap s(x_1,x_2)]\] The problem now is that this is equivalent to the formula for the extracted $vp$ we computed before!

Though it would seem that there is not much of a difference between concatenating the empty string to the left or to the right of an $np$ constituent, there should be a difference in behaviour between an $np\backslash s$ gap and a $s/np$ gap: for example, the first, but not the second can be modified by an subject-oriented adverb of type $(np\backslash s)\backslash(np\backslash s)$. The naive first-order translation fails to make this distinction.

There is a solution, and it consists of moving the universal quantifier out. Instead of the universal quantifier having only the $A$ formula as its scope, we turn it into an existential quantifier which has the entire $A\multimap C$ formula as its scope as follows.
\begin{align*}
\| A \multimap C \|^{y,z} & = \exists x. [\| A \|^{x,x} \multimap \| C
\|^{y,z} ] 
\end{align*}
 This allows us to correctly distinguish these two cases, but at the price of no longer having a residuated pair for the extraction phenomena\footnote{This analysis also makes an unexpected empirical claim: the treatment of parasitic gapping in type-logical grammars using the linear logic exponential $!$ would require the exponential to have scope over the quantified variable representing the empty string. We therefore need to claim that parasitic gapping can only happen with atomic formulas.}.


\editout{
\begin{align*}
\exists v. [a(x,y) \otimes b(v,v)] &\vdash c(x,y)\\
a(x,y)\otimes b(v,v) &\vdash c(x,y) \\ 
a(x,y) &\vdash b(v,v) \multimap c(x,y) \\
a(x,y) &\vdash \exists V. [b(V,V) \multimap c(x,y)] \\	
\end{align*}

\begin{align*}
a(x,y) &\vdash \exists V.[b(V,V) \multimap c(x,y)] \\	
a(x,y) &\vdash b(v,v) \multimap c(x,y) \\
a(x,y)\otimes b(v,v) &\vdash c(x,y) \\ 
\exists v. [a(x,y) \otimes b(v,v)] &\vdash c(x,y)\\
\end{align*}
}

\section{Discussion}
\label{sec:discussion}

One obvious aspect of first-order linear logic which hasn't been mention thus far is that the Horn clause fragment corresponds to a lexicalised version of multiple context-free grammars \parencite{wijnholds,moot13lambek}. Horn clauses for first-order linear logic are of the form $\forall x_0,\ldots,x_n [ p_1 \otimes \ldots \otimes p_m \multimap q]$ for predicates $p_i$ and $q$, or equivalently $\forall x_1,\ldots,x_n.(p_1 \multimap ( \ldots \multimap (p_m \multimap q)$, and they code each segment of an MCFG by a pair of string positions. In the context of MCFG it is well-known that each additional segment increases the generative capacity. When the maximum arity is 2, each predicate has a single segment and we have context-free grammars allowing us to generate languages such as $a^nb^n$. When the maximum arity is 4, we can generate $a^nb^nc^nd^n$, with maximum arity 6 $a^nb^nc^nd^ne^nf^n$, and so on \parencite{kallmeyer}. 

It is unclear which of these classes best captures the properties we want with respect to the string languages needed for the analysis of natural languages. It is generally assumed that a reasonable minimum is 4 (that is, two string segments per predicate). For example the languages generated by tree adjoining grammars and several similar formalisms are strictly included in this class (more precisely, the tree adjoining languages have the additional contraint of well-nestedness, whereas the multiple context free languages in general do not \parencite{mcfg}).

It is unclear to me which would be the right number of components to consider. Values between 4 and 6 components would seem to suffice for most applications, and it is unclear whether there are good linguistic reasons for abandoning well-nestnedness. 

The well-nested, residuated connectives seem to be the same as those definable in the Displacement calculus. Indeed, I have elsewhere already implicitly assumed a linear order for all subproofs when relating the Displacment calculus to first-order linear logic \parencite{moot13lambek}.

One interesting area of further investigation would be to relax the linear order constraint. For example, we let our sequent compute a unique partial order over the initial position variables (now no longer linearly ordered) and consider the sentence grammatical when the input string is a valid linearisation of this partial order. This would be potentially interesting for languages with relatively free word order.

\section{Conclusions}

This paper has discussed several aspect of adding partial order constraints to first-order linear logic. 
Although somewhat odd from the logical point of view, adding order constraints to the variables in first-order linear logic allows us to preserve the standard algebraic and category theoretic perspectives on type-logical grammars. In addition, some linguistically interesting operations can only be defined as part of a residuated triple when we impose partial order constraints on the string position variables. 

We have also shown how partial order constraints can be use as a mechanism for improving proof search by filtering out choices inconsistent with this order.





\printbibliography
\end{document}